\documentclass{article}
\usepackage[utf8]{inputenc}
\usepackage{authblk}
\usepackage{setspace}
\usepackage[margin=1.25in]{geometry}
\usepackage{graphicx}
\usepackage{makecell}
\graphicspath{ {./figures/} }
\usepackage{subcaption}
\usepackage{caption}
\usepackage{amsmath}
\usepackage{bm}
\usepackage{mathtools}
\usepackage{float}

\usepackage[style=ieee, 
citestyle=numeric-comp,
sorting=none]{biblatex}
\title{Machine learning approaches for feature engineering of the crystal structure: Application to the prediction of the formation energy of cubic compounds}

\author[1]{Prathik R. Kaundinya}
\author[2,3]{Kamal Choudhary}
\author[1,4*]{Surya R. Kalidindi}

\affil[1]{School of Computational Science and Engineering, Georgia Institute of Technology, Atlanta, Georgia 30332, USA.}
\affil[2]{Materials Science and Engineering Division, National Institute of Standards and Technology, Gaithersburg, Maryland 20899, USA.}
\affil[3]{Theiss Research, La Jolla, CA 92037, USA.}
\affil[4]{G.W. Woodruff School of Mechanical Engineering, Georgia Institute of Technology, Atlanta, Georgia 30332, USA.}
\affil[*]{Corresponding author. Email: surya.kalidindi@me.gatech.edu}

\date{May 23, 2021}

\begin{document}
	
	\maketitle
	
	\begin{abstract}
		In this study, we present a novel approach along with the needed computational strategies for efficient and scalable feature engineering of the crystal structure in compounds of different chemical compositions. This approach utilizes a versatile and extensible framework for the quantification of a three-dimensional (3-D) voxelized crystal structure in the form of 2-point spatial correlations of multiple atomic attributes and performs principal component analysis to extract the low-dimensional features that could be used to build surrogate models for material properties of interest. An application of the proposed feature engineering framework is demonstrated on a case study involving the prediction of the formation energies of crystalline compounds using two vastly different surrogate model building strategies - local Gaussian process regression and neural networks. Specifically, it is shown that the top 25 features (i.e., principal component scores) identified by the proposed framework serve as good regressors for the formation energy of the crystalline substance for both model building strategies.
	\end{abstract}
	
	
	\section{Introduction}
	Although physics-based modeling approaches such as density-functional theory (DFT) \cite{RN1,RN2} offer the preferred avenue for estimating the physical and chemical properties of crystal structures, they are not ideally suited for materials discovery and innovation efforts. Such efforts demand inverse solutions such as finding the molecular chemistry and structure that meets a targeted combination of physical and chemical properties \cite{RN3}. One of the most practical strategies for addressing the inverse solutions of materials design is to first produce high-fidelity low-computational cost surrogate models trained to the available data (e.g., collections of DFT computations), and then use the surrogate model for addressing the inverse problems of interest. Such a strategy can prove to be highly beneficial in rapid screening of vast design spaces \cite{RN3,RN4,RN5}. In this context, the emerging toolsets and paradigms in Machine Learning (ML) offer new opportunities for building the surrogate models needed to accelerate the discovery of new crystal structures. In particular, it is now possible to train these models using collections of publicly accessible DFT data in repositories such as the Materials Project (MP), Open Quantum Materials Database (OQMD), Joint Automated Repository for Various Integrated Simulations (JARVIS), Computational 2D Materials Database (C2DB) and Quantum Machine (QM) \cite{RN6,RN7,RN8,RN9,RN10,RN11,RN12,RN13,RN14,RN15,RN16}.
	
	ML models have been used in prior work to predict a broad range of material properties based on DFT datasets. The properties modeled have included optical and electronic bandgaps \cite{RN17,RN18}, formation energy \cite{RN6,RN8,RN19,RN20,RN21,RN22,RN23,RN24}, atomization energies \cite{RN25,RN26,RN27,RN28,RN29} and polarizability of crystalline compounds \cite{RN30}. ML models have also been used in prior work for the prediction and classification of crystal compositions and structures \cite{RN31,RN32,RN33,RN34,RN35} and in the development of many-body interatomic potentials for atomistic simulations \cite{RN36,RN37,RN38}. One of the foundational elements of ML common to most model building approaches is feature engineering, which aims to identify a small list of transformed input variables (called features) from the original large list of input variables that potentially could influence the predictions of the output variables (called targets). As one would expect, feature engineering governs the accuracy and utility of the surrogate model. In some of the ML approaches (e.g., convolutional neural networks (CNNs)), feature engineering occurs implicitly in the model training phase. In general, the more implicit one makes the feature engineering task, the more training data would be needed. This is because the model needs to learn the salient features first before learning their quantitative relationships with the targets. In much of the prior work \cite{RN12,RN21,RN25,RN33}, large feature lists have been manually cultivated by researchers based largely on their intuition of the atomic physics. Consequently, these feature lists have included various simplified attributes of the chemical compositions (e.g., atomic fractions) as well as the atomic structure (e.g., bond lengths and bond angles). These approaches face significant challenges. First, the creation of these lists has been pursued largely in an ad-hoc manner, which inevitably reflects the bias of the individual researchers. As such, these approaches may fail to account for certain potentially salient features controlling the target property. Second, the specification of chemical composition using distinct labels for the different atomic species hinders learning across crystal structures with different elemental compositions. One strategy to address this challenge has been to supplement the feature lists with certain atomically weighted physical properties \cite{RN8,RN39,RN40}. Additionally, there have also been various efforts at enhancing the specification of the atomic structure in the feature lists by retaining much of the actual three-dimensional (3-D) atomic structure information. For instance, Faber et al. \cite{RN20} utilized an extension of the Coulomb matrix representation to account for the periodicity of the crystal structure. In a different approach, Ward et al. \cite{RN23} represented the crystal structure with a Voronoi decomposition and employed local atomic species-level descriptors between neighboring species in addition to global structure descriptors such as maximum packing efficiency. In similar efforts, Schutt et al. and Honrao et al. used a feature set containing values of the partial Radial Distribution Function (RDF) between pairs of constituent species in the crystal structure \cite{RN41,RN42}. The partial RDF is a measure of the frequency of a pair of species separated by a certain distance within the material volume. However, the partial RDF does not contain directional information about the distribution of the atoms. One approach to overcome this limitation has been to supplement the partial RDF with an Angular Distribution Function (ADF) that contains information on the distribution of inter-atomic angles between species \cite{RN43}. Nevertheless, this approach still utilizes distinct labels for constituent species, and is thus typically useful only for description of crystal structures with few species (such as Al-Ni and Cd-Te systems \cite{RN43}).  Another interesting approach proposed recently relied on the development of graph embeddings that encode species-specific features (such as the periodic table group number, period number, electronegativity) and a limited set of structural features (such as the interatomic distance) as nodes and edges of a multigraph, respectively \cite{RN24,RN44,RN45}. This representation is amenable to usage in graph convolutional neural networks \cite{RN46,RN47} that sequentially build up localized representations for each node by iteratively including information from neighboring nodes.
	
	Using the feature engineering schemes described above, current efforts have employed various regression approaches such as kernel ridge regression (KRR) \cite{RN20,RN42}, support vector machines \cite{RN42,RN43}, deep neural networks \cite{RN21}, gradient boosted decision trees \cite{RN8}, and graph convolutional neural networks \cite{RN24,RN44,RN45} to build the desired reduced-order models. While the regression-based learning methods are computationally very efficient, they are often prone to over-fitting because they invariably employ a large number of implicit model parameters that need to be trained on a large collection of ground-truth data. The overfitting may be mitigated to a certain extent with the adoption of well-known techniques such as early stopping \cite{RN48}, loss function regularization and dropout \cite{RN49,RN50}. With the adoption of such techniques, the high computational efficiency of neural networks (NNs) makes them an attractive option for building reduced-order models. However, a significant challenge arises from the availability of relatively small training datasets due to the very high cost of DFT computations. The role of the feature engineering procedure is especially critical in such smaller sized datasets because of the need to restrict the number of model fit parameters that can be used in these cases.
	
	ML approaches based on Bayesian inference offer an alternate option that is likely to prove beneficial for building surrogate models from DFT computations. More specifically, Gaussian process regression (GPR) \cite{RN51} offers a powerful non-parametric Bayesian approach to building surrogate models from small training datasets. There are many potential benefits to the utilization of Gaussian Processes in the context of regression. First, they allow a formal treatment of uncertainty in the model predictions. In other words, they do not just estimate the expected values for the model outputs, but also their distributions. Second, the formal treatment of model uncertainty allows the design and implementation of strategies that could potentially reduce the effort spent in the generation of the training data. More specifically, it is possible to formulate and maximize the expected information gain to the surrogate model with the addition of each specific new training data point. This is particularly important to building surrogate models with limited data from the computationally expensive DFT datasets. Furthermore, GPR models are typically more interpretable than NNs. 
	
	In this paper, we present a systematic and comprehensive approach to feature engineering of crystal structure that directly addresses the challenges described earlier. In this new approach, we systematically and efficiently assemble an extremely large number of spatial correlations (specifically, 2-point correlations) \cite{RN5,RN52} that implicitly account for atomic attributes (e.g., Pauling electronegativity, ionization energy, heat of fusion). After processing the large feature list for the complete ensemble of crystal structures of interest, we employ principal component analysis (PCA) to obtain a suitable low-dimensional representation of the feature list. This strategy offers many advantages compared to the approaches used in current literature. First, the approach presented in this work has the potential to systematically generate a very large set of physics-inspired features for the rigorous and comprehensive quantification of the crystal structure. Second, the proposed protocols utilize digital (i.e., voxelized) representations of the 3-D crystal structure \cite{RN53,RN54} along with compact Fourier representations and associated computational algorithms (e.g., Fast Fourier transform (FFT)) for highly efficient computation of the features (i.e., spatial correlations). Third, the encoding of individual chemical species through their individual physical properties enables effective learning across crystal structures of different elemental compositions. Fourth, the use of PCA provides an objective (i.e., data-driven) path for establishing low-dimensional representations of the compound crystal structure that can then be used with a broad variety of model-building approaches. Fifth, the features identified by the proposed framework are independent of the target property predicted by the model. 
	
	The primary goal of this paper is to develop and demonstrate a versatile feature engineering methodology for materials problems involving different chemical compositions of compounds and their crystal structures. This methodology is aimed at extracting reliable models from small data sets using physics-inspired feature engineering approaches. We demonstrate the utility of this feature engineering scheme by building predictive models for the crystal formation energy using two drastically different model-building approaches: (i) a localized variant of GPR \cite{RN55}, and (ii) a feed-forward neural network (NN). A second goal of this work is to critically compare the relative merits of both model building techniques for addressing materials problems.
	
	\section{Background}
	The central goal of the feature engineering step is to establish a compact set of salient inputs (i.e., features) that serve as suitable predictors for the selected targets. In the context of our problem, this would entail establishing a set of salient crystal structure descriptors for capturing high-fidelity reduced-order Structure-Property (SP) relationships \cite{RN5,RN56}. In this effort, the formalism of n-point spatial correlations (also called n-point statistics) \cite{RN5,RN57,RN58,RN59,RN60,RN61} offers a systematic approach to statistical quantification of the underlying morphological patterns in the heterogeneous material internal structure. Several existing statistical atomic physics models such as the Ising model (for predicting ferromagnetism) \cite{RN62} and the Potts model (a general model of interacting atomic spins) \cite{RN63} predict the bulk properties of materials as a sum of contributions arising from local structure features that can be easily interpreted as components of the n-point spatial correlations. Spatial correlations are indeed the features dictated by the governing physics in studies of heterogeneous material structure at all length scales, spanning from the atomistic to the mesoscale. However, the comprehensive set of n-point spatial correlations is typically too large and unwieldy to serve directly as inputs for the reduced-order models. As such, there is a critical need for a compact representation of the material structure features that can serve effectively as inputs to produce the desired high-fidelity reduced-order models. In this section, we present a brief overview of the concepts of spatial correlations, salient feature extraction (using PCA), and the strategies for building reduced-order models (i.e., GPR and NNs) used in this work.
	
	\subsection{Spatial Correlations}
	Typically, efficient computations of spatial correlations for the quantification of the heterogeneity of the material structure utilize digital representations (i.e., voxelized representations with suitable assignments of material states to each voxel) and FFT algorithms for computing the convolution operations involved in these computations \cite{RN5}. Mathematically, these voxelized representations can be expressed as a high-dimensional array $m_s^p$, whose elements capture the value of a local feature (reflecting the local material state) indexed by p in a voxel indexed by s (defined as vector of integers $\{s_1,s_2,s_3\}$ for 3-D material volumes). This array of size $(P\times n(S))$, with P denoting the number of distinct material states assigned to each voxel in the material volume and $n(S)$ denoting the cardinality of the set of spatial voxels $S$, defines one instantiation of a material structure. 
	
	Our interest lies in extracting microstructure statistics from each instantiation (i.e., $m_s^p$ array) that can serve as features in the formulation of surrogate models connecting material structure to the properties/performance characteristics of interest. The most systematic set of such microstructure statistics are provided by the formalism of n-point spatial correlations, which capture the relevant statistics related to the morphological details (i.e., size and shape distributions) of the distinct local states in the material structure. The most basic set of these statistics are the 2-point correlations denoted by $f_r^{pq}$, which track how often distinct local states $p$ and $q$ are separated by a discretized vector indexed by $r$. Mathematically, 2-point correlations may be computed as \cite{RN5,RN61,RN64}
	\begin{equation} \label{eq:1}
		f_r^{pq} = \frac{\sum_{\mathbf{s}\in S}m_\mathbf{s}^pm_{\mathbf{s+r}}^q}{n(S)}
	\end{equation}
	
	One way to interpret the above definition is to recognize that the numerator denotes the number of successes in locating the local states $p$ and $q$ separated by the vector index $r$, and that the denominator represents the total number of valid trials (for periodic microstructures, this is equal to the number of voxels). It should be noted that the RDF (used commonly in the quantification of the molecular structures) \cite{RN59,RN65} is a particular variant of the 2-point correlations described in Eq. (1) in which one does not consider the orientation of the vector, but only its magnitude. Noting that the central operation in Eq. (1) is essentially a discrete cross-correlation over the spatial domain of voxels $S$, the 2-point correlations are efficiently computed by taking advantage of FFT algorithms \cite{RN5,RN52}. The computations naturally exploit the periodicity of the crystal structures. There also exist several redundancies that can be leveraged for obtaining a more compact set of spatial correlations. Specifically, it can be seen that Eq. (1) produces $P^2$ sets of spatial correlations. It has been shown that only P of these are adequate to compute the rest of the spatial correlations \cite{RN61,RN66}. As a result, it is often adequate to compute the autocorrelations (which correspond to $n=p$ in Eq. (1)) for the dominant local state in the collected ensemble of material structures and its cross-correlations with the rest of the local states.
	
	\subsection{Extraction of salient features}
	In prior work, our research group has established a generalized framework referred as Materials Knowledge Systems (MKS) \cite{RN67,RN68,RN69} that has demonstrated the versatility and utility of assembling a large set of n-point spatial correlations and then applying PCA to establish suitable data-driven low-dimensional representations of the material internal structure. PCA offers a dimensionality reduction technique that performs a rotational transformation of the features into an orthogonal space organized to maximize the capture of variance in the given dataset. The transformed axes (i.e., new basis) are known as PCs, while the transformed coordinates are known as PC scores. However, since PCA maximizes the capture of variance, it tends to emphasize the features that exhibit the largest variance in the dataset. Therefore, one could apply suitable scaling factors to various subgroups of the features before performing the PCA, in order to adjust their roles (i.e., increase or decrease their importance in the PCs). In this work, the features are scaled such that the total variance in the spatial correlations corresponding to each selected pair of local states (i.e., each combination of $p$ and $q$) is the same. This type of scaling equalizes the importance of each subset of spatial correlations related to a specific combination of local states in the PCA. This protocol produces an ordered list of transformed features such that the selection of each additional transformed feature maximizes the capture of the explained variance in the dataset. A truncated list of these ordered transformed features (i.e., PC scores) serve as suitable low dimensional representations of the material structure. 
	
	The benefits of the MKS pipeline have been expounded in prior work \cite{RN5,RN56,RN69,RN70,RN71,RN72,RN73}. First, this approach yields an objective low-dimensional representation of the material structure that is not influenced by either the materials manufacturing process or property information, thereby providing a consistent representation of the material structure in process-structure-property (PSP) linkages \cite{RN5}. Second, since PCA is essentially a Fourier representation, it allows for approximate but optimal reconstructions of the material structure statistics (controlled by the truncation levels applied in retaining the PC scores), which in turn can then be used with sophisticated algorithms \cite{RN74,RN75} for the statistical reconstruction of the material structure. Third, the concepts based on spatial correlations and PCA are broadly applicable to virtually all different classes of material internal structures found at multiple hierarchical length scales, spanning from the atomic to the macroscale. Fourth, and perhaps the most importantly, it has been seen that only a handful of PC scores ($\approx 5$ to $10$) are often adequate in producing high fidelity PSP linkages needed to drive materials innovation efforts \cite{RN5,RN70,RN76}. It should be noted that the use of the PC representations of the spatial correlations as the features of the material structure has thus far been explored mostly at the meso-length scales \cite{RN5,RN70,RN73,RN77}. This concept is only now beginning to be explored at the atomic structure length scales.
	
	\subsection{Reduced-order model building strategies}
\begin{figure}[H]
		\centering
		\includegraphics[width=0.8\textwidth]{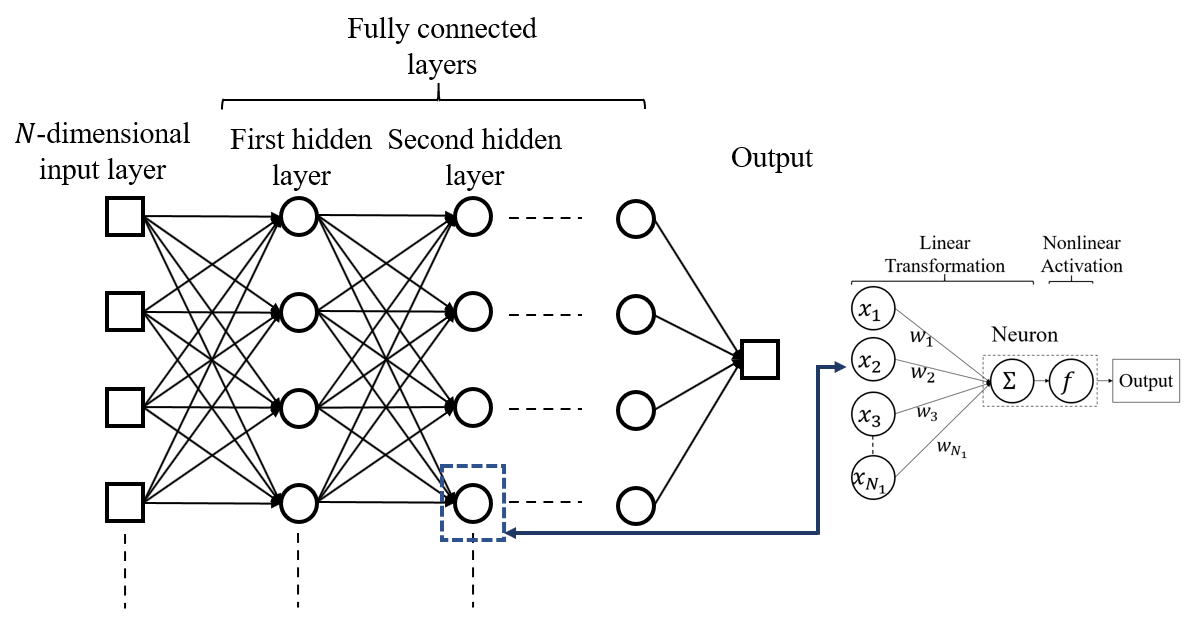}
		\caption{Schematic description of a fully connected neural network for an $N$-dimensional input. The callout describes the details of the computations performed at each neuron.}
		\label{fig:1}
	\end{figure}

	In the final step of the MKS framework, the reduced-order features of interest are correlated with the target property using a variety of model-building strategies. NNs provide a nonlinear modeling approach to accomplish this task and are known to be sufficient to capture any arbitrary mapping between the inputs and the output. The most basic feedforward NNs consist of multiple fully connected layers of multiple neurons, with each neuron capturing a linear transformation followed by a nonlinear activation (e.g., ReLu, sigmoid function) \cite{RN78}. The architecture of a typical feed-forward NN is shown in Fig. \ref{fig:1}. The weights and the bias (i.e., model fitting parameters) associated with each neuron are calibrated with backpropagation. This is accomplished through minimization of a user-specified loss function between the predicted output and the corresponding observed values in the training data. NNs derive their scalability from the computationally efficient algorithms used for the calibration of the model parameters. These are readily accessible in many software packages (e.g., PyTorch \cite{RN79}, TensorFlow \cite{RN80}), most of which are amenable to computation on Graphics Processing Units (GPUs). NNs also typically iterate through multiple epochs (i.e., passes over the entire training data set) in order to optimize the model parameters. The main limitation of the NNs is that they need a substantially large training dataset, without which they are most likely to produce a model overfit (i.e., significantly larger errors for test data points compared to the training data points). Overfits to training data often occur due to an over-parametrization of the NN (i.e., more trainable weights and biases to be calibrated than the size of the dataset). In this work, we utilize feedforward NNs with two hidden layers to predict the formation energies of the compounds in our dataset. For an NN with two hidden layers consisting of $h_1$ and $h_2$ neurons, respectively, the number of trainable model parameters to be calibrated (denoted by $C$) for a single output from an N-dimensional input is given by
	\begin{equation} \label{eq:2}
		C=(Nh_1+h_1h_2+h_2)+(h_1+h_2+1)
	\end{equation}
	
	An alternate modeling technique that is better suited for small datasets is GPR \cite{RN51,RN81,RN82}, which offers a nonparametric Bayesian approach to building surrogate models. GPR (and its variants) have successfully been applied to model structure-property linkages in the mesoscale from relatively small datasets \cite{RN70,RN83,RN84}. GPR models the target as a Gaussian Process (GP) that is fully defined by the specification of a mean and a covariance. The GP is then tuned using conditional distributions defined on available training data. Let $\bm{X}$, $\bm{X}^*$ and $\bm{y}$ denote the $N\times D$ matrix of training data points, $N^*\times D$ matrix of test data points and $N\times1$ vector of target values in the training set, respectively. Additionally, let $\bm{K(X,X^{'})}$, $\bm{K(X^*,X^{*'})}$ and $\bm{K(X,X^{*'})}$ denote the $N\times N$, $N^*\times N^*$ and  $N \times N^*$ covariance matrices assembled (using a kernel function described later) using the inputs in the training dataset, test dataset, and between the training and test datasets, respectively. In GPR, the predictive mean and variance for test points is expressed as
	\begin{equation} \label{eq:3}
		\bm{\mu^*=K(X,X^{*'})^\top K(X,X^{'})^{-1}y}
	\end{equation}
	\begin{equation} \label{eq:4}
		\bm{\Sigma^*=K(X^*,X^{*'})-K(X,X^{*'})^\top K(X,X^{'})^{-1}K(X,X^{*'})}
	\end{equation}
	where $\top$ denotes the transpose of a matrix. As stated earlier, one typically uses a suitable kernel function to compute the various covariance matrices in Eqs. \ref{eq:3} and \ref{eq:4}. One of the most commonly used kernel functions is the ARD-SE (Automatic Relevance Determination – Squared Exponential) \cite{RN51,RN85} function expressed as
	\begin{equation} \label{eq:5}
		k(\bm{x,x^{'}})=\sigma_s^2 \sum_{d=1}^D exp\left(\frac{-(x_d-x_d^{'})}{2\theta_d^2}  \right)+\sigma_n^2\delta_{\bm{xx^{'}}}
	\end{equation}
	where $\bm{x}$ and $\bm{x^{'}}$ denote any two input vectors selected from the data matrices for which the covariance matrix is being computed, subscript d corresponds to the $d^\text{th}$ feature in the input vector, $\sigma_s$ is a scaling parameter that controls the scaling of the output variance, $\theta_d$ is the correlation-length hyperparameter corresponding to the $d^\text{th}$ feature in the input vector, $\sigma_n$ denotes the noise in the target, and $\delta_{\bm{xx^{'}}}$ denotes the Kronecker delta. The treatment of the noise using a single parameter added to the diagonal of the covariance indicates independence of noise with the input data (i.e., homoscedasticity) \cite{RN51}. The central benefit of the ARD-SE kernel is that it allows us to tune independently the correlation-length hyperparameter for each of the input features. This allows for better interpretability of the model, because the salient features with relatively smaller length scale hyperparameters exhibit a higher sensitivity to the predicted target. One should also note that extremely small values of the hyperparameters tend to make the predictions very noisy. Therefore, it is important to tune the values of the hyperparameters in order to produce robust and reliable predictions. This is typically accomplished with Maximum Likelihood Estimation (MLE), which lacks a closed-form solution; consequently, one must resort to iterative schemes such as quasi-Newton algorithms to optimize the hyperparameters \cite{RN86,RN87,RN88}.
	
	One of the central challenges in the implementation of GPR comes from the computational cost, especially with large training datasets. The main computational bottleneck arises from the computation of the inverse of $\bm{K(X,X^{'})}$ in Eq. \ref{eq:3}, which typically scales as $\mathcal{O}(N^3)$ for dense matrices. This makes traditional GPR intractable even for moderately sized datasets $(N>1,000)$. This difficulty is compounded by the need to optimize the large number of hyperparameters $(D+2)$ in the ARD-SE kernel. Several methods have been suggested in literature to address these challenges. These include the utilization of low-rank sparse approximations for the covariance matrix \cite{RN51,RN89,RN90}, tree-based partitioning to develop smaller datasets and fit individual GP predictors \cite{RN91}, and Localized GPR (L-GPR) proposed by Gramacy and Apley (discussed next and used in this study) \cite{RN92}. 
	
	L-GPR involves identifying a local subset of training data points to construct a separate GP for each predictive point. In addition to allowing the application of GPR to larger datasets, a key advantage of L-GPR is the ability to accommodate non-stationarity in the model by allowing optimization of the interpolation hyper-parameters in the kernel functions depending on the location of the test point in the high-dimensional input space. Different criteria may be considered while selecting the local training points; an intuitive but suboptimal method would be using a certain number of nearest neighbors as the local neighborhood \cite{RN93}. However, a more informative design criterion would be to sequentially build the local neighborhood by evaluating a tradeoff between expected information gain by adding a training point and the increase in the predictive variance of the GPR. This method of local training point selection promotes an efficient exploration of the sample space and provides a natural guidance to avoid overfitting of the models. This approach is referred as Active Learning Cohn (ALC) \cite{RN55,RN94}, and is utilized in this work.
	
	\section{3. Novel framework for quantifying crystal structures}
	In this section, we describe a new framework for effective low-dimensional representation (i.e., fingerprints) of the crystal structure of the compounds. This is accomplished by first establishing computationally efficient protocols to arrive at voxelized representations of the atomic structure of compounds, and then suitably extending and applying the current MKS framework (described in Sections 2.1 and 2.2) on such voxelized representations. In prior work \cite{RN74,RN95}, very simple descriptors were utilized for the microstructure array, $m_{\bm{s}}^n$, where it was assigned a value of one for voxels within the atomic volume and a value of zero otherwise. A key challenge encountered in this simple representation is that it does not allow an efficient interpolation between material structures (i.e., compounds) of varying chemical compositions. In this work, we expand the previous framework to allow for an enhanced representation of multiple atomic attributes of interest in each voxel. This is accomplished by representing the local material state as a vector set of attributes that take continuous values. Such continuous local states have been utilized previously in the treatment of mesoscale polycrystalline microstructures and mesoscale fields of chemical compositions \cite{RN5,RN96}. In this work, we will extend these representations to the quantification of crystal structures of compounds. 
	
	Choudhary et al. \cite{RN8} have studied the order of importance of various atomic attributes and their role in the formation energy of a compound. Specifically, they have identified ionization energy, Pauling electronegativity and heat of fusion as the most important attributes. Therefore, in this work, we define the local material state in the voxel indexed by s using a vector of attributes $\langle a_{0s},a_{1s},a_{2s},v_s \rangle$, where $a_{0s}$, $a_{1s}$ and $a_{2s}$  denote the heat of fusion, ionization energy, and Pauling electronegativity, respectively, and $v$ is a binary attribute that takes a value of zero for voxels within any of the atomic volumes (defined by a sphere with radius equal to the atomic radius of the species) and a value one otherwise. Our model building efforts (described later) found it beneficial to further enhance the attribute list using various monomials of the atomic attributes. In other words, the full list of local states included in the atomic state vector $\bm{m_s}$  may be defined as
	
	\begin{equation} \label{eq:6}
		\bm{m_s}=\langle a_{0s},a_{1s},a_{2s},a_{0s}a_{1s},a_{1s}a_{2s},a_{0s}a_{2s},a_{0s}a_{1s}a_{2s},a^2_{0s},a^2_{0s}a_{1s},\ldots,a^{l_0}_{0s}a^{l_1}_{1s}a^{l_2}_{2s},v_s \rangle
	\end{equation}
	where $l_0$, $l_1$ and $l_2$ represent the maximum degree considered for each of the three atomic attributes. A total of $(l_0+1)\times (l_1+1)\times (l_2+1)$ local states were used in the definition of the atomic state vector. Through multiple trials, it was found that the combination of $l_0$, $l_1$ and $l_2$ set to $5$, $5$ and $2$, respectively, yielded the optimal performance in terms of the tradeoff between computational cost (which scales with the number of local states), and the accuracy of our models. For these selections, the total number of local states in $m_s$ is $108$. Elements of $\bm{m_s}$ are then used in the computation of the 2-point correlations using Eq. (1). As already noted, the convolution operation involved is best accomplished using the FFT algorithm. Mathematically, this is expressed as 
	\begin{equation} \label{eq:7}
		F_{\bm{k}}^{pq}=\frac{1}{n(S)}M_{\bm{k}}^{p*}M_{\bm{k}}^{q}
	\end{equation}
	where $M_k^p$ and $F_k^{pq}$ denote the discrete Fourier transforms of $m_s^p$ and $f_r^{pq}$, respectively, and the superscript * denotes the complex conjugate. It is important to note that Eq. (7) implicitly assumes the fields involved are spatially periodic. Although all the compounds considered in this work exhibit periodicity (i.e., all spatial fields defined based on the atomic structure of the compound are inherently spatially periodic), the sizes of their periodic unit cells vary substantially. For example, the lattice parameter (reflecting the length scale of the periodicity) for the compounds included in this study varies over the range of 1.8 Å to 12.4 Å. Therefore, if one were computing the spatial correlations using voxelization schemes based on the respective unit cell sizes, Eq. (7) can be applied directly without any problems. However, a constant voxel size is essential for obtaining the desired low-dimensional representations using PCA in the later steps. Since the unit cell size L of the different compounds considered in this study is unlikely to be an integer multiplier of a common voxel size, t, a suitable strategy is needed to efficiently compute the spatial correlations using a standardized voxel size. Given the size of the dataset, these computations are only practical if we can continue to exploit the cost savings provided by the FFT algorithm in evaluating the spatial correlations. A suitable computational protocol has been devised to address this challenge, which is described next. This new protocol builds on prior efforts involving mesoscale material microstructures \cite{RN5}.
	
\begin{figure}[H]
		\centering
		\begin{subfigure}[b]{0.33\textwidth}
			\centering
			\includegraphics[width=\textwidth]{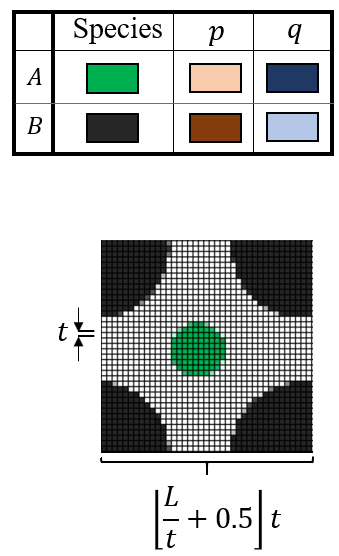}
			\caption{}
			\label{fig:2a}
		\end{subfigure}
		\hfill 
		\begin{subfigure}[b]{0.33\textwidth}
			\centering
			\includegraphics[width=\textwidth]{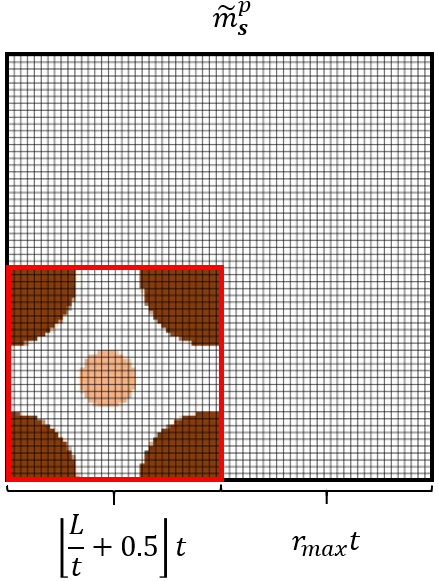}
			\caption{}
			\label{fig:2b}
		\end{subfigure}%
		\begin{subfigure}[b]{0.33\textwidth}
			\centering
			\includegraphics[width=\textwidth]{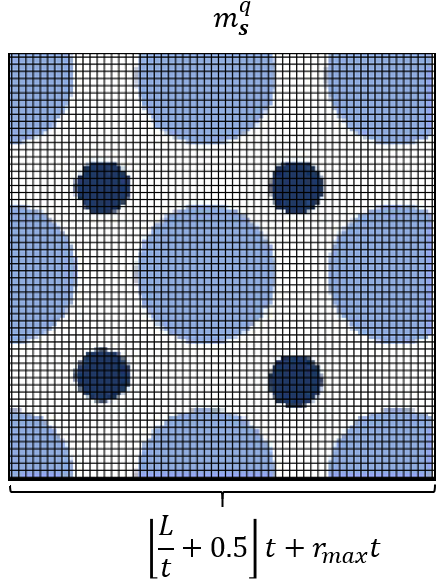}
			\caption{}
			\label{fig:2c}
		\end{subfigure}
		\caption{(a) A typical 2-D cross section of a unit cell of the compound structure, with green and black regions representing two different chemical species. (b) and (c) Extended compound structures depicting the spatial distribution of the local states $p$ and $q$, respectively. These extended structures are designed such that the spatial correlations to the desired vector length (i.e., $r_{max}$ in each of the vector components) can be computed efficiently using the FFT algorithms that implicitly treat the functions as being periodic. The red border in (b) identifies the outline of original unit cell from (a). } \label{fig:2}
	\end{figure}

	The computational scheme presented in this work is designed to use a single voxel size and a common set of corresponding vectors indexed by {r}, for the computation of sets of 2-point correlations (i.e., $f_r^{pq}$) using Eqs. (1) and (7) for all the compounds considered in this study. For 3-D volumes, it is convenient to represent r itself as a vector of integers $\langle r_1,r_2,r_3 \rangle$. The 2-point correlations used in this study are computed only for the set of positive vectors up to a selected maximum length for each of the three components, i.e., $\{r|0 \leq r_1 \leq r_{max},0 \leq r_2 \leq r_{max},0 \leq r_3 \leq r_{max} \}$. Therefore, the total number of discrete vectors for which we seek to compute $f_r^{pq}$ is $r_{max}^3$. Our approach to efficiently compute these correlations considers the numerator and denominator in Eq. (1) separately. As mentioned earlier, for each selected r, the numerator represents the number of successes in locating the local states p and q separated by the selected vector. Our strategy to accurately (and efficiently) compute this quantity is to express the numerator as a cross-correlation between two different material structure spatial fields, $\tilde{m} _{s}^p$ and $m_s^q$, both of which are derived from the original material structure. The construction of these structure fields is illustrated in Fig. \ref{fig:2} for a simple crystalline compound comprising two chemical species A and B. Note that all structure fields shown in Fig. \ref{fig:2} are 2-D sections of the 3-D material volume of this compound. Fig. \ref{fig:2a} shows the unit cell of the original compound structure within a volume that is closely approximated by an integer number of voxels of the selected size, t. Therefore, the voxelated structure shown in Fig. \ref{fig:2a} is no longer exactly periodic. In fact, the size of this structure is $\lfloor \frac{L}{t}+0.5\rfloor t$, where $\lfloor \cdot \rfloor$ indicates the greatest integer function (also referred to as the floor function). Clearly, the size of the voxelized unit cell shown in Fig. \ref{fig:2a} is not equal to L. This discrepancy introduces a small error in the computation of the spatial correlations that scales with the value of t. In other words, one would have to select the value of t carefully, as a compromise between minimizing the voxelization error and keeping the computational cost reasonable. 
	
	The structure fields $\tilde{m}_s^p$ and $m_s^q$ are defined over larger spatial domains as shown in Figs. \ref{fig:2b} and \ref{fig:2c}. The size of these fields is specified as $\lfloor \frac{L}{t}+0.5\rfloor t + r_{max}t$, while the corresponding set of voxels is labelled as $S^*$. Note that the fields are extended so that when any of the vectors of interest (identified earlier as $\{r\}$) are placed with their tails in the unit cell structure shown in Fig. \ref{fig:2a}, their heads are guaranteed to lie within the extended spatial domains shown in Figs. \ref{fig:2b} and \ref{fig:2c}. Furthermore, the two fields in these figures are associated with local states p and q identified in the definition of the subset of spatial correlations $f_r^{pq}$. The structure fields shown in Figs. \ref{fig:2b} and \ref{fig:2c} have been specifically designed so that the cross-correlation defined as $\sum_{\bm{s}\in S*}\tilde{m}_{\bm{s}}^p m_{\bm{s+r}}^q$ provides a sufficiently accurate estimate of the numerator to within the inherent voxelization error discussed earlier. This is because $\tilde{m}_s^p$ returns the value of the local state $p$ only when the tail of the vector indexed by r lies within the original unit cell shown in Fig. \ref{fig:2a}. This region is indicated by the red outline in Fig. \ref{fig:2b}. More importantly, this structure field returns a zero value otherwise, which nullifies any possible contributions arising from the wrap-around vectors implicit in the Fourier transform operations performed in Eq. (7). Note that this nullification only works for vectors within the selected set $\{r\}$. In other words, the use of Eq. (7) on the fields shown in Figs. \ref{fig:2b} and \ref{fig:2c} produces estimates of the numerator in Eq. (1)for many more vectors not included in $\{r\}$. In the strategy described here, one retains only the results for the vectors in $\{r\}$  and discards the rest of the results. Although one computes more values than needed in this protocol, the overall computational cost is much cheaper than the direct computation of the correlations exclusively for the vector set of interest. The denominator in Eq. (1) is simply the number of voxels in the original unit cell shown in Fig. \ref{fig:2a}, and is given by $(\lfloor\frac{L}{t}+0.5\rfloor)^3$. The protocol described here was found to provide an excellent approximation for the fast computation of the desired spatial correlations. In our work, we implemented the aforementioned protocol with the value of $t$ and $r_{max}$ selected as $0.2 Å$ and $30$, respectively, which corresponds to the computation (and retention) of the 2-point correlations for vectors with components up to $6Å$. This selection was motivated by the fact that 92 \% of the crystalline compounds present in our dataset had a lattice parameter less than 6 Å. 
	
\begin{figure}[H]
		\centering
		\begin{subfigure}[b]{0.33\textwidth}
			\centering
			\includegraphics[width=0.8\textwidth]{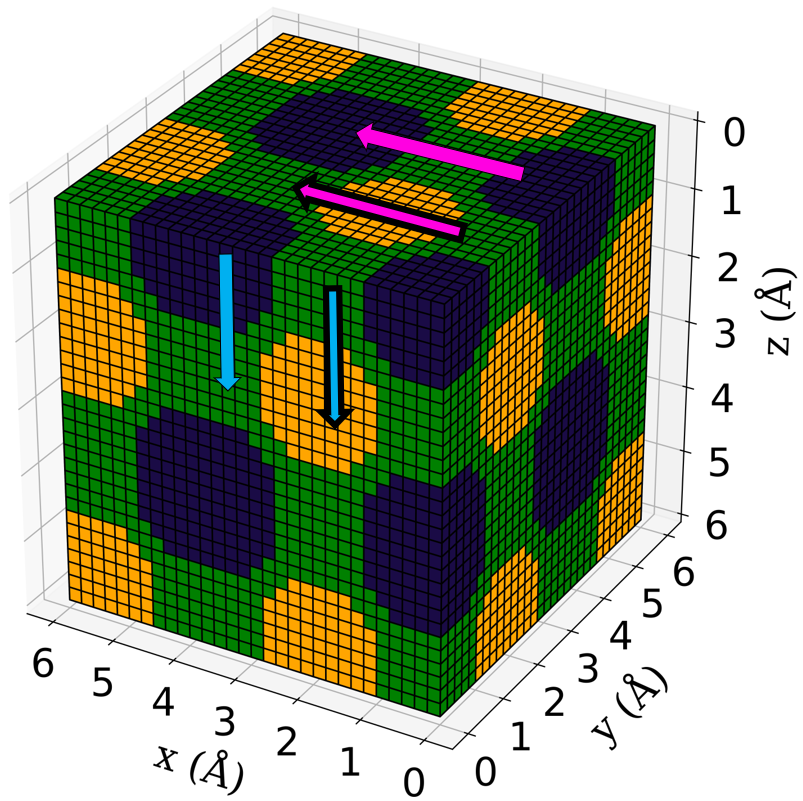}
			\caption{}
			\label{fig:3a}
		\end{subfigure}
		\hfill 
		\begin{subfigure}[b]{0.33\textwidth}
			\centering
			\includegraphics[width=0.9\textwidth]{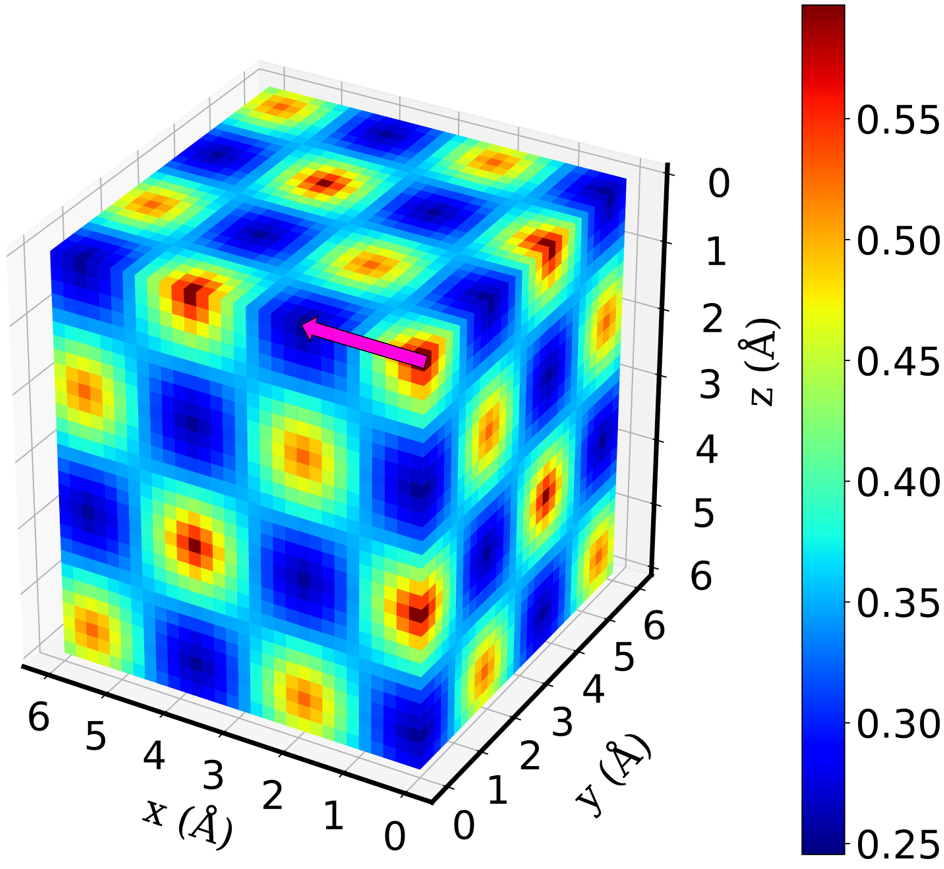}
			\caption{}
			\label{fig:3b}
		\end{subfigure}%
		\begin{subfigure}[b]{0.33\textwidth}
			\centering
			\includegraphics[width=0.9\textwidth]{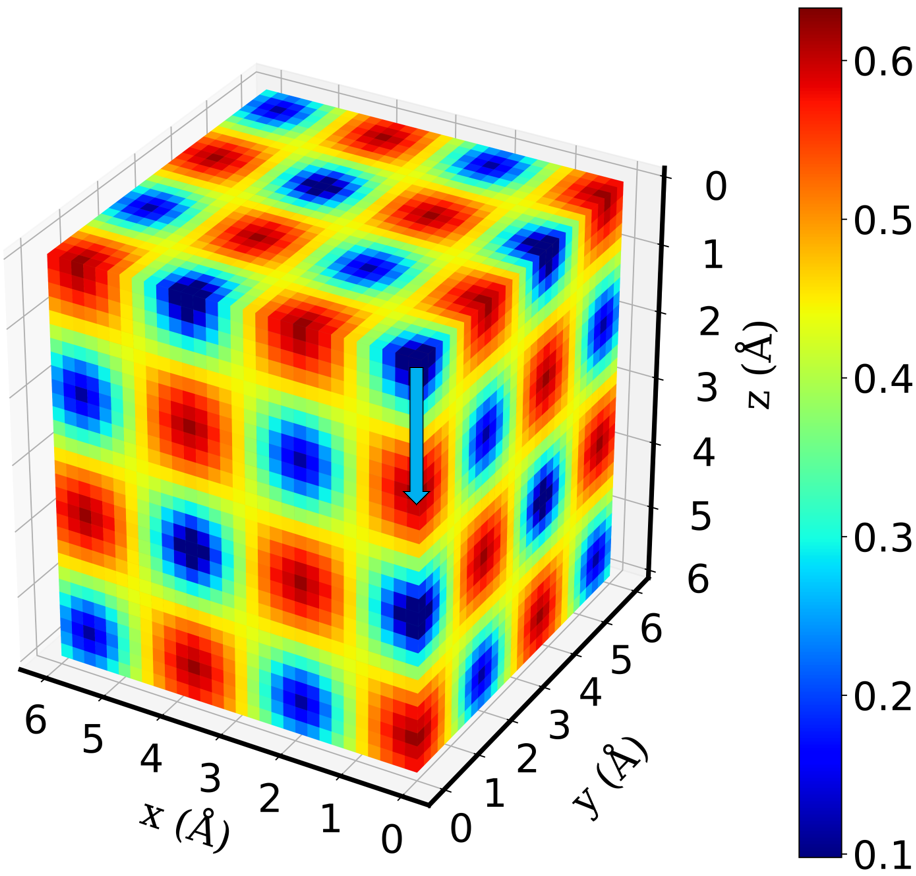}
			\caption{}
			\label{fig:3c}
		\end{subfigure}
		\caption{(a) Voxelized 3-D crystal structure of $AlNi_3$ with the green, orange, and dark blue colors representing the voxels occupied by the void, $Al$, and $Ni$ species. (b) The autocorrelation map for the void state $v_s$.  (c) The $v_s-a_{2s}$ cross-correlation map.} \label{fig:3}
	\end{figure}

	Figs. \ref{fig:3a}, \ref{fig:3b} and \ref{fig:3c} show an example crystal structure consisting of two atomic species $AlNi_3$, its autocorrelation map for $v_s$ (the void state), and its $v_s-a_{2s}$ cross-correlation map, respectively. Note that the auto- and cross-correlation maps shown correspond only to a single octant in the vector space (with positive values of $r_1$, $r_2$ and $r_3$). It can be observed that the periodicity of the crystal structure is indeed captured in these maps. The value of the autocorrelation corresponding to the zero vector ($0.58$ in Fig. \ref{fig:3b}) reflects the void volume fraction in the crystal structure. Since the void state in the present application is set to either zero or one in each voxel, the autocorrelation map in Fig. \ref{fig:3a} actually provides statistical information for a large set of vectors, whose heads can be selected anywhere in the depicted vector space, but tails fixed at $\langle 0,0,0 \rangle$. As a specific example, the pink-colored vector in Fig. \ref{fig:3b} corresponds to $\langle 3.8,0,0 \rangle Å$  , with a void autocorrelation of $0.58$. This autocorrelation value simply reflects the probability of finding two void (green) voxels in the crystal structure shown in Fig. \ref{fig:3a} separated by the selected vector. Fig. \ref{fig:3a} shows two example placements of the selected vector – a successful placement (pink vector with a black outline) connecting two void (green) voxels and an unsuccessful placement (pink vector without a black outline).  Due to the periodicity of the crystal structure, it is seen that the autocorrelation value for the pink-colored vector is the same as autocorrelation for the zero vector (i.e., the void volume fraction). The cross-correlation map shown in Fig. \ref{fig:3c} similarly captures the spatial correlations between states $v_s$ and $a_2s$ present in the structure. Note that since we do not allow the void state and any material state to co-exist in a single voxel, the cross-correlation for the zero vector exhibits a zero value. A cross-correlation peak is observed for $\langle 0,0,1.8 \rangle Å$ in Fig. \ref{fig:3c}, shown using a blue-colored vector. As before, two example placements of this vector are shown in Fig. \ref{fig:3a}, with one reflecting a successful sampling of the desired spatial correlation.
	
	\section{Dataset}
	A dataset of compound crystal structures and formation energies was obtained from the publicly accessible JARVIS-DFT repository \cite{RN7,RN9,RN10,RN11,RN12}. The extracted data included the crystal structure information (i.e., Bravais lattice type and the coordinates of the atoms present in the unit cell) and their DFT-computed chemical properties. Specifically, the formation energies of $\sim 1740$ cubic crystalline compounds that were computed with the OptB88vdW functional \cite{RN97} were extracted from this repository, and were used to train our models and evaluate the utility of our feature engineering scheme. The compounds included in this dataset comprised 70 different chemical species, thereby offering opportunities for the extraction of surrogate models applicable to a broad range of chemical compositions. The most frequently occurring non-metallic species in the dataset were oxygen and nitrogen, while the most frequently appearing metallic species were aluminum, palladium and rubidium. The least frequently appearing metal atoms are cesium, rhenium and technetium. Among the crystalline compounds in the dataset, there are 1160 intermetallic compounds, 408 metal oxides, 95 metal halides, and 37 metal sulfides. Moreover, the dataset consisted of 21, 950, 743 and 27 unary, binary, ternary and quaternary crystalline compounds. 
	
\begin{figure}[H]
		\centering
		\begin{subfigure}[b]{0.42\textwidth}
			\centering
			\includegraphics[width=\textwidth]{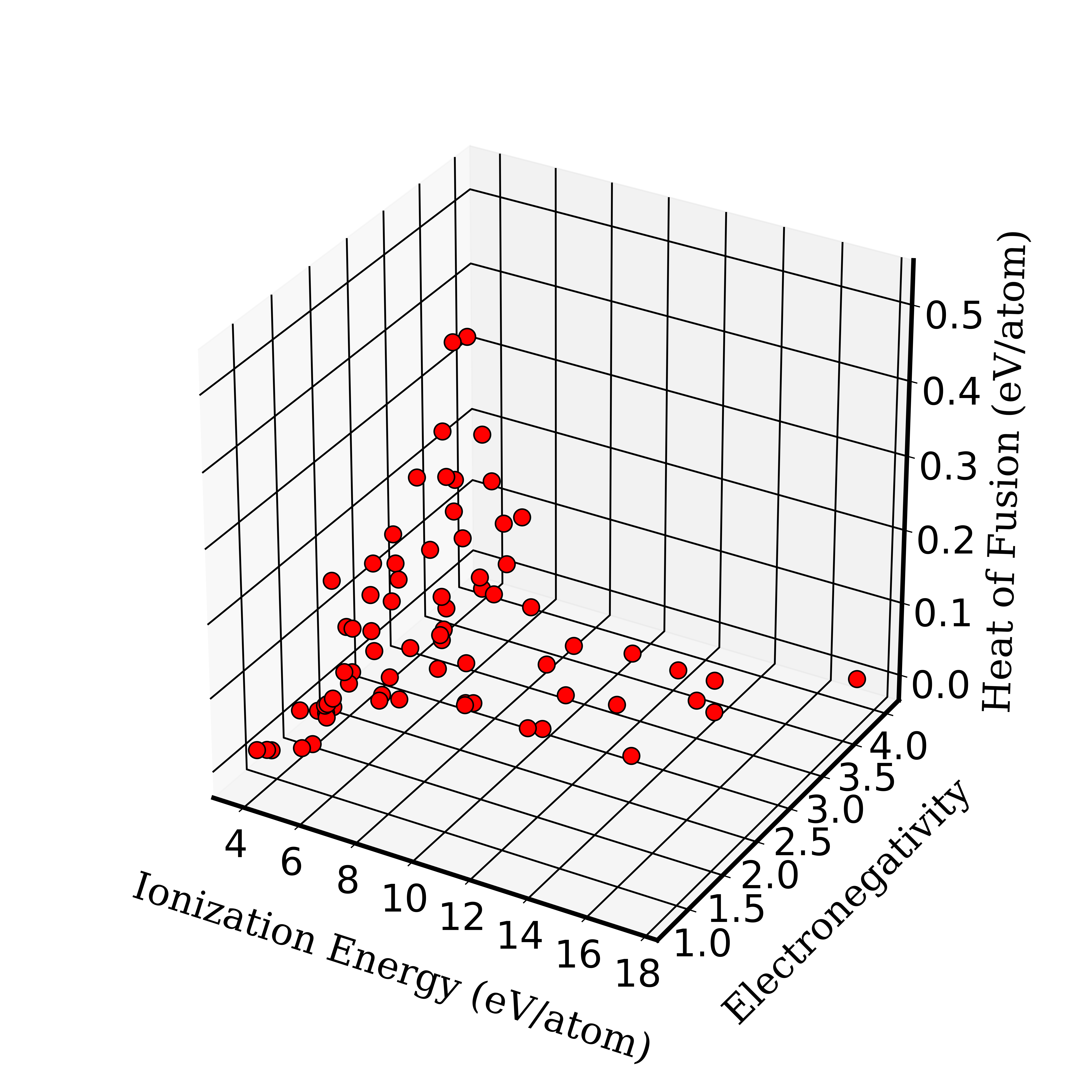}
			\caption{}
			\label{fig:4a}
		\end{subfigure}
		\hfill 
		\begin{subfigure}[b]{0.5\textwidth}
			\centering
			\includegraphics[width=\textwidth]{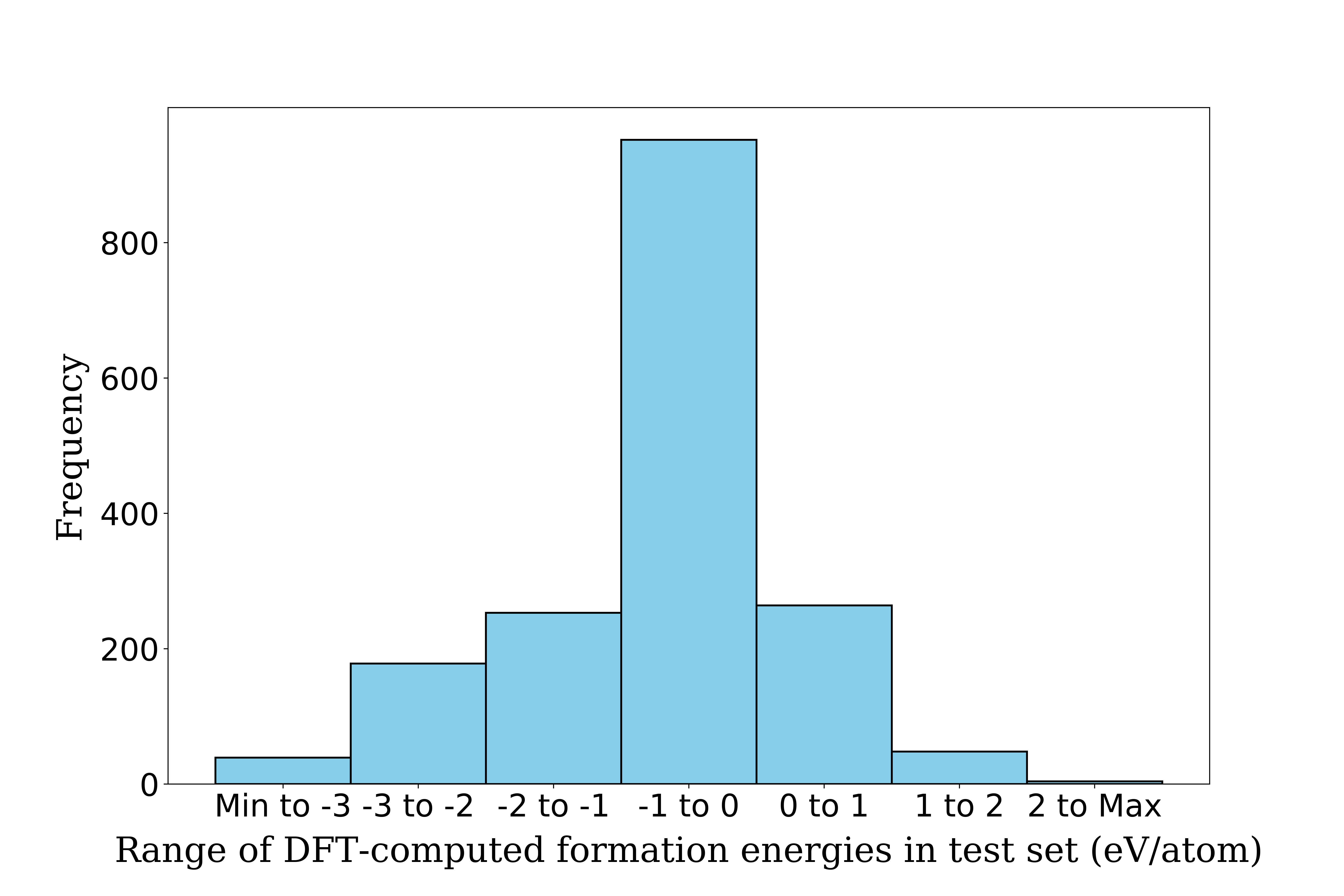}
			\caption{}
			\label{fig:4b}
		\end{subfigure}
		\caption{(a) Distribution of the attribute values for the different atomic species present in the dataset. (b) Histogram of the formation energies of the compounds in the dataset.} \label{fig:4}
	\end{figure}

	Fig. \ref{fig:4a} shows the distribution of the values of the three local state attributes (i.e., heat of fusion ($a_0$)  ionization energy ($a_1$), Pauling electronegativity ($a_2$) for all of the different chemical species present in the dataset. The electronegativity of the different chemical species in the dataset appears to be fairly evenly distributed over the range 0.79 to 3.98. In contrast, the distributions of the heat of fusion and ionization energy appear to be non-uniformly distributed, with more of the chemical species having attribute values in the lower end of their respective ranges of $(2.3e-3 \text{ to } 0.52)$ eV/atom and $(3.89 \text{ to } 17.42)$ eV/atom. Fig. \ref{fig:4b} represents the distribution of the DFT-computed formation energies of compounds (the target or output for the models) present in the dataset. Note that the samples are unevenly distributed in the values of the formation energy, with most of the samples having a formation energy between -3 and 1 eV/atom. From this distribution of target properties, one would expect a higher accuracy in the prediction of the formation energy within this region, and a lower accuracy outside it.
	
	The feature engineering framework developed in Section 3 is applied to the selected dataset. As described in section 2.1, the complete set of $P^2$ (or $108^2$) spatial statistics contains several redundancies. Building on prior work [refs], the following sets of spatial correlations were included in this work: (i) autocorrelations of $v_s$, $a_{0s}$ and $a_{1s}$  (this were the most dominant signals in the dataset), and (ii) the cross-correlations of these three local states with the rest of the local states, while eliminating the trivially related sets (the $f_r^{pq}$ are trivially related to $f_r^{qp}$). This resulted in a total of 321 sets of spatial correlations for each crystal compound. The dimensionality reduction procedure described in section 2.2 was then applied on the complete set of spatial statistics to obtain 25 PC scores for each compound. Since each PC basis in our application carries (weighted) information from a total of $\sim3$ million 2-point statistics, its precise interpretation is currently impractical. 
	
	\section{Results and discussion}
	As mentioned earlier, two different modeling strategies have been explored in this study to evaluate critically the efficacy our feature engineering framework. These modeling strategies were specifically chosen to represent distinctly different approaches to building surrogate models. Therefore, our main goal here is to evaluate the performance of the proposed feature engineering framework (involving PCA on a large feature vector of spatial correlations that account rigorously for the atomic details in the crystal structure) on distinctly different model building strategies. The selected model building strategies for this study include L-GPR (implemented using the laGP package in the R programming language \cite{RN55}) and NN (implemented using PyTorch \cite{RN79}). The inputs and the output to both models were the set of the top 25 PC scores (scaled to unit variance) generated using the feature engineering framework presented in this work and the DFT-computed formation energy, respectively. Both modeling strategies benefitted significantly from the scaling of inputs (to unit variance), due to their usage of gradient-based optimization in the regression procedure (i.e., hyperparameter tuning in L-GPR, and loss function minimization in NNs). 
	
	In this study, Mean Absolute Error (MAE) and Median Absolute Error (MedAE) have been chosen as error metrics for quantifying the predictive accuracy of the surrogate models. For a set of $J$ predictive test samples, these error measures are expressed as
	\begin{equation} \label{eq:8}
		MAE=\frac{1}{J}\sum_{j\in J}|y^{(j)}_{pred}-y^{(j)}_{actual}|
	\end{equation}
	
	\begin{equation} \label{eq:9}
		MedAE=Median(|y^{(1)}_{pred}-y^{(1)}_{actual}|,|y^{(2)}_{pred}-y^{(2)}_{actual}|,\ldots,|y^{(J)}_{pred}-y^{(J)}_{actual}|)
	\end{equation}
	in which $y^{(j)}_{actual}$ and $y^{(j)}_{pred}$ indicate the DFT-computed and the surrogate model predicted values of the formation energy for the sample j, respectively. While the same error metrics are used to evaluate performance of both modeling strategies, the training and test protocols are substantially different because of the very different underlying philosophies involved in these model building strategies. These are described next.
	
	\subsection{L-GPR modeling approach}
\begin{figure}[H]
		\centering
		\begin{subfigure}[b]{0.29\textwidth}
			\centering
			\includegraphics[width=\textwidth]{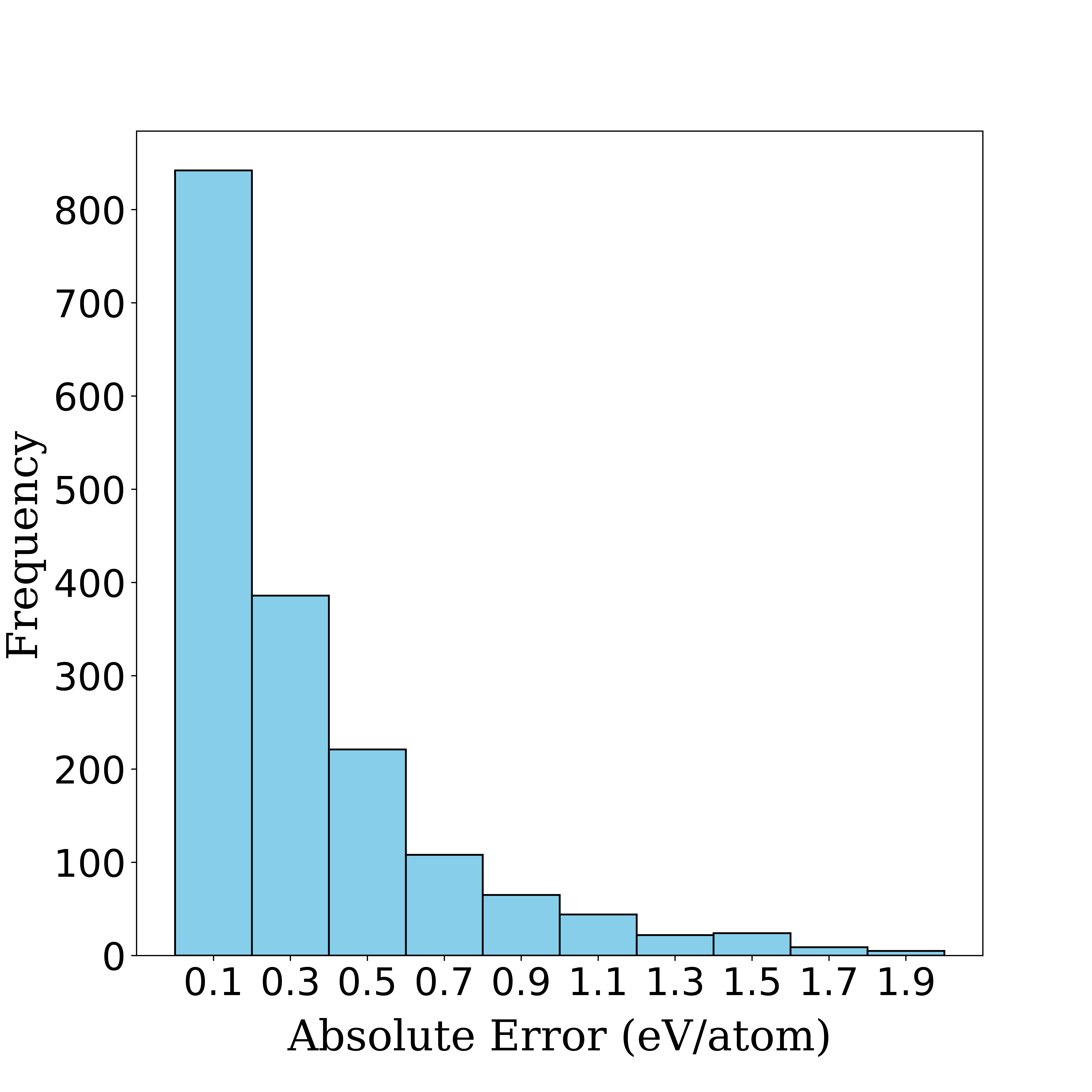}
			\caption{}
			\label{fig:5a}
		\end{subfigure}
		\hfill 
		\begin{subfigure}[b]{0.29\textwidth}
			\centering
			\includegraphics[width=\textwidth]{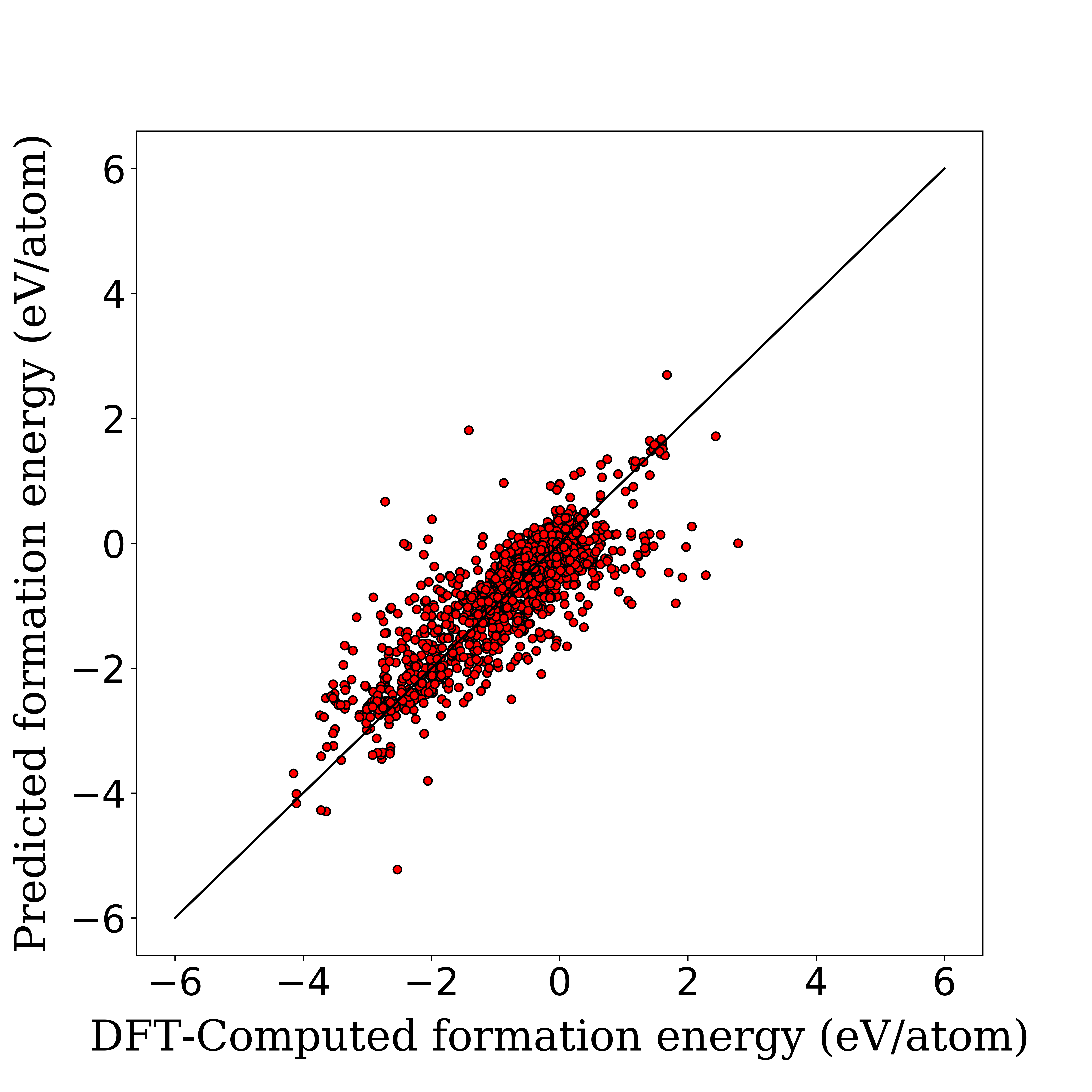}
			\caption{}
			\label{fig:5b}
		\end{subfigure}
		\begin{subfigure}[b]{0.4\textwidth}
			\centering
			\includegraphics[width=\textwidth]{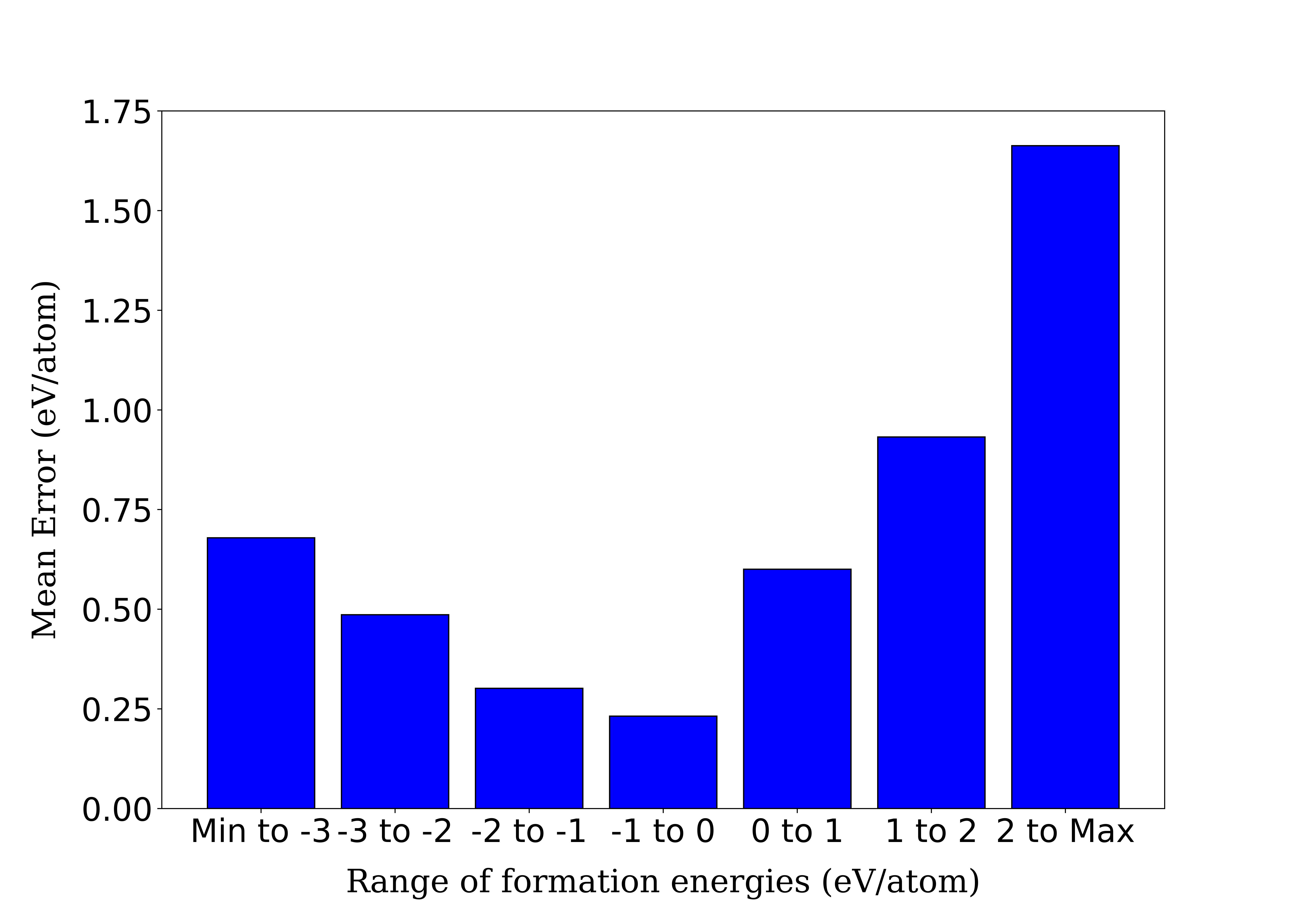}
			\caption{}
			\label{fig:5c}
		\end{subfigure}
		\caption{(a) Histogram of errors obtained in prediction of the DFT-computed formation energy for L-GPR Model 4 in Table \ref{tab:1} with the local training dataset size $\chi$ of 200 points. (b) Parity plot of the L-GPR predicted vs DFT-computed formation energies for the same L-GPR model for all data points used in this study. (c) Bar plot of the mean prediction error (of the same model) in the formation energy for different ranges of DFT-computed formation energies.} \label{fig:5}
	\end{figure}

	In this study, five different models were built using the L-GPR strategy. Recall that in this strategy, a separate GP is formulated for each point in the dataset using training points selected from the local neighborhood based on the ALC criterion described in Section 2.3. Since predictions were made at each point independently, the $MAE$ and $MedAE$ were computed directly with Eqs. (9) and (10) over all samples (indexed by $j$) in the dataset (consisting of $J$ samples). These values are denoted as $MAE_{L-GPR}$ and $MedAE_{L-GPR}$, respectively. The main difference in the five L-GPR models built for this study was in the number of points selected from the local neighborhood (denoted by $\chi$). Table \ref{tab:1} summarizes the accuracies of these models, as evaluated by the error metrics described previously. In this table, the value of $\chi$ is also shown as a percentage of the total number of points in the dataset. The main purpose of this analysis was to understand the effectiveness of the ALC criterion in the L-GPR scheme for selecting optimal training points for the desired model. Since GPR is essentially an interpolation strategy, L-GPR offers significant computational savings compared to the conventional GPR strategy (which utilizes all points in the dataset). This is because the computational cost of the GPR scales as $\mathcal{O}(N^3)$, as mentioned previously. Also, the use of ALC in the L-GPR to select the training points offers an organic strategy for regularization and avoiding over-fit. More specifically, since the ALC criterion selects neighborhood points based on the goal of maximizing information gain, one would expect that after the training dataset has included most of the “important” points, there would only be an incremental improvement in performance with a further increase in $\chi$. This tendency was indeed observed in our results. Table \ref{tab:1} shows that a selection of about 200 points in the neighborhood of each test point in the 25-dimensional input space (i.e., with $\chi=200$) provides a robust prediction of the formation energy. Fig. \ref{fig:5a} shows the histogram of errors of the formation energies predicted by this model. As seen in this figure, around 80 \% of the formation energies are predicted to within an error of 0.4 eV/atom. Only a small improvement in overall predictive performance was observed when $\chi$ was further increased to 250 points.
	
	
	\begin{table}[H]
		\renewcommand{\arraystretch}{1.2}
		\centering
		\begin{tabular}{|p{1.5cm}|p{1cm}|p{2cm}|p{2.5cm}|p{2.5cm}|}
			\hline
			\centering{Model No.} & \centering{$\chi$} &  \centering{Percentage of entire dataset (\%)} & \makecell[t]{$MAE_{L-GPR}$ \\ (eV/atom)} &  \makecell[t]{$MedAE_{L-GPR}$ \\ (eV/atom)}\\ \hline\hline
			\hfil LGPR1 &  \hfil 50 &  \hfil 2.84 &  \hfil 0.393 &  \hfil 0.256\\ 
			\hfil LGPR2 &  \hfil 100 &  \hfil 5.68 &  \hfil 0.378 &  \hfil 0.244\\ 
			\hfil LGPR3 &  \hfil 150  &  \hfil 8.52 &  \hfil 0.352 &  \hfil 0.219\\ 
			\hfil LGPR4 &  \hfil 200 &  \hfil 11.36 &  \hfil 0.330 &  \hfil 0.192\\ 
			\hfil LGPR5 &  \hfil 250 &  \hfil 14.20 &  \hfil 0.329 &  \hfil 0.187\\ 
			\hline
		\end{tabular}
		\caption{The predictive accuracy of the five L-GPR models built in this work to estimate the DFT-computed formation energies. The input to all models is the set of the 25 PC scores of the spatial correlations of the crystal structure. }
		\label{tab:1}
	\end{table}
	
	\noindent Fig. \ref{fig:5b} shows the parity plot of the L-GPR predicted vs DFT-computed formation energies corresponding to Model 4 from Table \ref{tab:1} (i.e., $\chi$=200). Fig. \ref{fig:5c} presents a bar plot of the distribution of the average error for the different values of the formation energies. As seen from the bar plot, the average error in the predicted values is lowest for samples with DFT-computed formation energies in the range of (-2 to 0) eV/atom. This range is also the region with the highest concentration of training data points (see Fig. \ref{fig:4b}). As expected, the accuracy of the L-GPR models built is quite sensitive to the amount of available training data close to test data points. For example, the average error for the samples with DFT-computed formation energies greater than 2 eV/atom is quite high, as there are only 20 data points with the corresponding formation energies. The predictive uncertainty of the L-GPR model also followed a similar distribution over these ranges of formation energies, with the highest uncertainty associated with the crystals with DFT-computed formation energies greater than 2 eV/atom. This indicates that prediction confidence in regions with more data is significantly higher than in the regions with sparse data.
	
	\subsection{NN modeling approach}
	Five different feedforward NN models (all consisting of two hidden layers) were evaluated in this study. For each model, the loss function (to be minimized) was chosen as L1 loss \cite{RN98}, which reflects the mean absolute error in the prediction of the target for the training dataset. This minimization was performed using the Adam gradient-based optimizer \cite{RN99} with a learning rate set to $5e-5$. The available dataset was partitioned into separate training and test datasets in a 80\% to 20\% split. This partitioning was done while maintaining similar distributions of the output values in the train and test datasets (similar to Fig. \ref{fig:4b}). A small fraction of the training dataset (about 5\%) was designated as a validation set, and utilized to implement an early stopping criterion \cite{RN48}, designed to mitigate over-fitting of the NN models produced. The performance of the model on the validation dataset is used to make decisions on when to stop the minimization iterations (epochs) on the loss function. This is especially important in the present work because of the relatively small training datasets, and relatively large number of trainable model parameters inherent to the NN models. The MAE and MedAE error metrics (denoted by MAENN and MedAENN, respectively) were computed using Eqs. (8) and (9) on the test data points for all NN models built in this study to evaluate their predictive accuracy. 
	
	Table \ref{tab:2} summarizes the accuracies of the five different NN models produced in this study. The numbers of neurons in the hidden layers were varied across all the NN models that were built in this study, in order to explore their influence on model performance. The number of the trained model parameters in each NN model (calculated using Eq.(2)) are also shown in Table \ref{tab:2}. It is seen that the predictive accuracy of the model does not increase significantly beyond NN3 model, in spite of large increases in the number of trainable model parameters. Indeed, with the number of trainable model parameters approaching the number of training data points, there is clear evidence of model overfit, especially with models NN4 and NN5. Fig. \ref{fig:6} shows a parity plot of the train and test predictions from the NN3 model. It can be seen in this plot that the prediction quality on the train and test sets are similar (with MAEs of 0.28 eV/atom and 0.34 eV/atom on each of these sets, respectively), indicating that the early stopping criterion provides an effective regularization method to help mitigate an overfit to the training set. Out of the set of compounds exhibiting a high prediction error (greater than 1 eV/atom), 28\% are metal oxides, 11\% are metal halides and 35\% are intermetallic compounds. Further analysis of the crystals having the ten highest prediction errors indicates that the prediction quality is affected by the proportion of occurrence of the atoms in the dataset. In each case, at least one of the constituent atoms occurs in less than 3\% (i.e., fewer than 52 samples) from the dataset. 
	
	
	\begin{table}[H]
		\renewcommand{\arraystretch}{1.2}
		\centering
		\begin{tabular}{|p{1.5cm}|p{2.5cm}|p{2cm}|p{2.5cm}|p{2.5cm}|}
			\hline
			\centering{Model No.} & \centering{Number of hidden neurons} &  \centering{Number of parameters} & \makecell[t]{$MAE_{NN}$ \\ (eV/atom)} &  \makecell[t]{$MedAE_{NN}$ \\ (eV/atom)}\\ \hline\hline
			\hfil NN1 &  \hfil (20,10) &  \hfil 741 &  \hfil 0.361 &  \hfil 0.232\\ 
			\hfil NN2 &  \hfil (20,14) &  \hfil 829 &  \hfil 0.352 &  \hfil 0.225\\ 
			\hfil NN3 &  \hfil (20,20) &  \hfil 961 &  \hfil 0.341 &  \hfil 0.198\\ 
			\hfil NN4 &  \hfil (25,20) &  \hfil 1191 &  \hfil 0.340 &  \hfil 0.201\\ 
			\hfil NN5 &  \hfil (25,25) &  \hfil 1326 &  \hfil 0.350 &  \hfil 0.195\\ 
			\hline
		\end{tabular}
		\caption{The predictive accuracy of the five NN models built in this work to estimate the DFT-computed formation energies. The input to all models is the set of the 25 PC scores of the spatial correlations of the crystal structure. Note that the error metrics ($MAE_{NN}$, $MedAE_{NN}$) were computed only for the $\sim345$ data points in the test set (represented by the green points in Figure \ref{Fig:6}).}
		\label{tab:2}
	\end{table}
	
	\begin{figure}[H]
		\centering
		\includegraphics[width=0.5\textwidth]{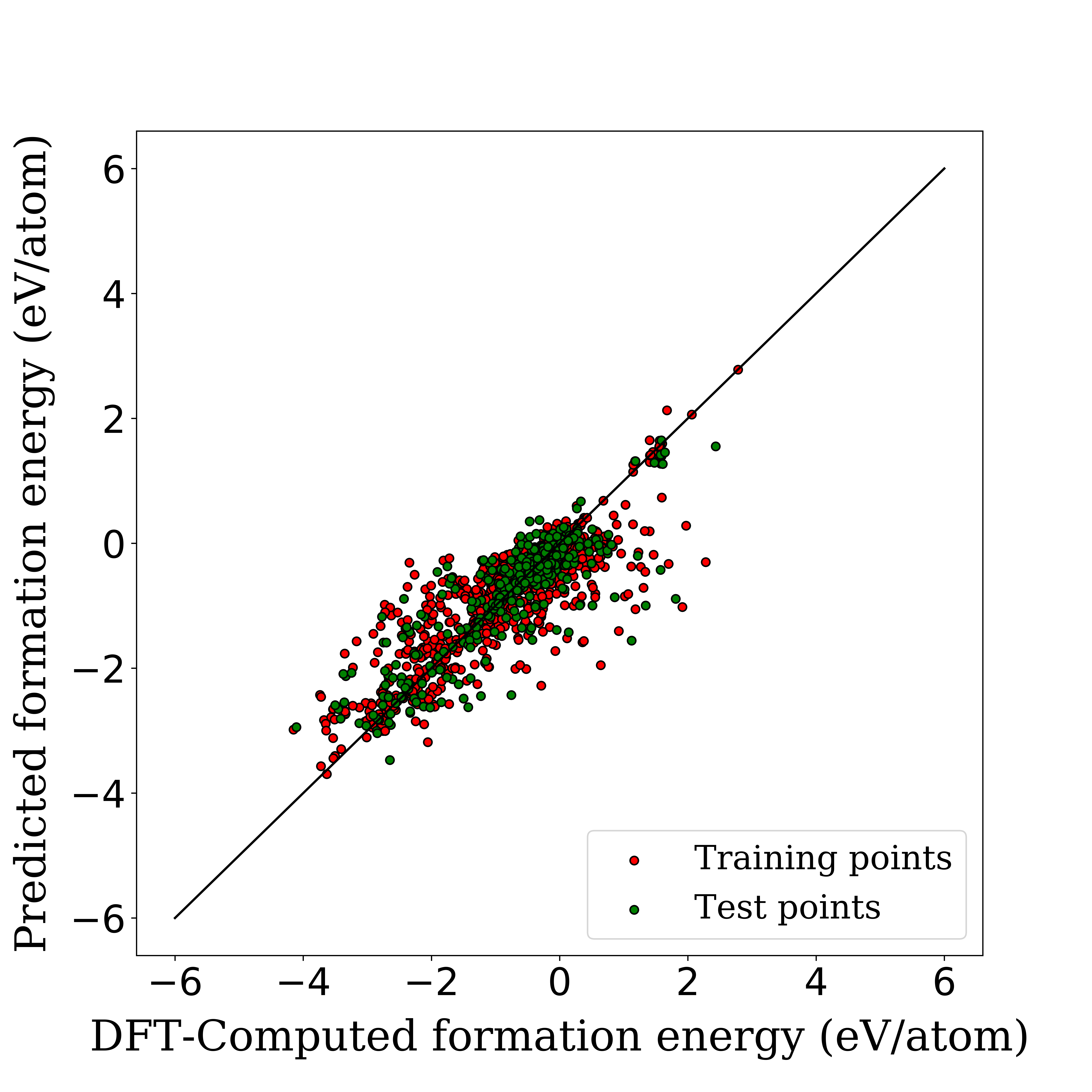}
		\caption{Parity plot of the NN-predicted vs DFT-computed formation energies for NN3 Model (see Table \ref{tab:2}). }
		\label{Fig:6}
	\end{figure}
	
	\subsection{Comparison of the L-GPR and NN surrogate model building strategies}
	While the accuracy of both modeling strategies explored in this study are reported using the same metrics (MAE and MedAE), the performance of these models are not easily compared directly because of the different underlying philosophies involved in the two strategies. However, they clearly point to the efficacy of the feature engineering paradigm presented in this work. It is indeed remarkable that the 25 features identified by our protocols (from an extremely large initial set of approximately three million 2-point correlations) provided reasonably accurate models when used as inputs to two completely different modeling strategies (i.e., the L-GPR strategy utilizing Bayesian local interpolations and the NN providing a global regression). This observation strongly supports our claim that the protocols involving spatial correlations and PCA are capable of identifying salient features that exhibit high utility in the formulation of surrogate structure-property models, independent of the modeling strategy employed. 
	
	The results from the models produced in this work are quite reasonable when compared with prior modeling efforts in literature. Faber et al. \cite{RN20} utilized three different structure representation schemes (Ewald sum matrix, extended Coulomb matrix, and Sine matrix) together  with a KRR modeling scheme and reported test set MAEs of 0.49 eV/atom, 0.64 eV/atom and 0.37 eV/atom, respectively, for the predicted values of formation energies. However, the authors also reported significantly lower training set MAEs of the order of 0.005 eV/atom for each of these models, indicating that the models obtained in this study are likely overfit to the training data. These models were trained on 3000 data points taken from the MP database \cite{RN16}. In a different study, Choudhary et al. [8] curated a large set of ~1550 chemical and structural descriptors as input features for establishing a model using a much larger dataset consisting of $\sim24,500$ materials. This study reported a test set MAE of 0.12 eV/atom with the use of an ensemble-based gradient-boosting decision tree (GBDT) regression model consisting of $\sim1150$ estimators with up to 270 leaves per estimator and an unconstrained tree depth (which was decided based on an early stopping criterion). This correlated with the use of $\sim500$ nodes for almost all estimators built, indicating that a large number of parameters were employed in the final model. While the test MAE reported in this study is lower, the feature engineering procedure used in this study is significantly more computationally intensive and difficult to scale to larger datasets. 
	
	It is emphasized that in comparison with the prior studies, the strategy developed in this work relies on unsupervised (i.e., independent of the selected output variable) feature engineering to identify a small number of salient features (PC scores). These features are then used as inputs to models with relatively fewer fit parameters.  This confirms the value and utility of the feature engineering protocols presented in this work. Both models produced in this study are likely to find uses in potentially different applications. The main strength of the L-GPR strategy is that it can provide objective guidance on what new training data points should be generated to improve the model fidelity. This is because the GPR models not only predict the expected values of the output for test data points, but also their variance. This attribute can be used in a suitable design of experiments strategy \cite{RN51,RN100} to identify which specific inputs exhibit the highest potential for model improvement. This is particularly useful in the initial stages of generating the training data sets, especially in situations where the cost of data generation is high (e.g., data generated using DFT models). In contrast, NNs could become the preferred approach after a substantial amount of data has been collected. This is mainly because of their ability to cover a much richer space of model functions, and their superior computational efficiency in handling large data sets. The data set used here is small enough that the L-GPR strategy is likely to be the better option for its size. 
	
	The overall predictive accuracy obtained by the models in this work is not sufficient for the accurate computation of the crystal formation energy. We believe that the model accuracy can be improved by (i) implementing the feature engineering scheme on a larger dataset (by lifting the constraint of cubic crystal symmetry), and (ii) using the obtained spatial correlations directly as inputs to other model architectures like CNNs. Larger training datasets would allow for models with a larger number of model fit parameters, and thereby improve the model accuracy. 
	
	\section{Conclusion}
	In this study, we have introduced a novel systematic and computationally efficient feature engineering framework based on 2-point spatial correlations for the quantification of cubic crystal structures with varied chemistries. The approach presented in this work employed a voxelized representation of the crystal structure and computed the spatial correlations on suitably selected atomic attributes (here taken as heat of fusion, ionization energy, and Pauling electronegativity). The proposed approach offers a novel avenue for feature engineering that will allow interpolations across different chemistries of the compounds. This work demonstrates that the proposed feature engineering approach combined with PCA for dimensionality reduction is capable of generating a compact set of salient features (i.e., PC scores) representing the many details of the crystal structure. These PC scores can be used effectively in building surrogate models needed to screen for materials exhibiting potential for improved properties. In this work, this was demonstrated by using the same features as inputs to two very different model-building strategies (i.e., L-GPR and NN) for the predictions of crystal formation energy. The results of this work indicate that the utility of the generated features is independent of the model-building strategy. Moreover, the distribution of errors in predictions (and their associated uncertainties in case of Bayesian approaches) offer objective guidance on where additional training data should be targeted to further improve the performance of the surrogate models. This work also found that the L-GPR modeling strategy produces more robust predictions when dealing with relatively smaller datasets, as the one utilized in this study. 
	
	\section*{Declaration of Competing Interest}
	The authors declare that no known competing financial interests have influenced the work reported in this paper.
	
	\section*{Acknowledgments}
	KP and SK gratefully acknowledge support from ONR N00014-18-1-2879. The Hive cluster at Georgia Institute of Technology (supported by NSF 1828187) was used for this work. The dataset used in this study is available to download from https://materialsdata.nist.gov/handle/11256/994 \cite{RN101}. 
	
	\printbibliography

@inproceedings{RN80,
author = {Abadi, Martin and Barham, Paul and Chen, Jianmin and Chen, Zhifeng and Davis, Andy and Dean, Jeffrey and Devin, Matthieu and Ghemawat, Sanjay and Irving, Geoffrey and Isard, Michael and Kudlur, Manjunath and Levenberg, Josh and Monga, Rajat and Moore, Sherry and Murray, Derek G. and Steiner, Benoit and Tucker, Paul and Vasudevan, Vijay and Warden, Pete and Wicke, Martin and Yu, Yuan and Zheng, Xiaoqiang},
title = {TensorFlow: A System for Large-Scale Machine Learning},
year = {2016},
isbn = {9781931971331},
publisher = {USENIX Association},
address = {USA},
pages = {265–283},
numpages = {19},
location = {Savannah, GA, USA},
series = {OSDI'16}
}

@article{RN4,
   author = {Agrawal, Ankit and Choudhary, Alok},
   title = {Perspective: Materials informatics and big data: Realization of the “fourth paradigm” of science in materials science},
   journal = {APL Materials},
   volume = {4},
   number = {5},
   DOI = {10.1063/1.4946894},
   url = {https://aip.scitation.org/doi/abs/10.1063/1.4946894},
   year = {2016},
   type = {Journal Article}
}

@inproceedings{RN22,
   author = {Agrawal, A. and Meredig, B. and Wolverton, C. and Choudhary, A.},
   title = {A Formation Energy Predictor for Crystalline Materials Using Ensemble Data Mining},
   booktitle = {2016 IEEE 16th International Conference on Data Mining Workshops},
   pages = {1276-1279},
   ISBN = {2375-9259},
   DOI = {10.1109/ICDMW.2016.0183},
   type = {Conference Proceedings}
}

@article{RN38,
   author = {Behler, Jörg},
   title = {Perspective: Machine learning potentials for atomistic simulations},
   journal = {The Journal of Chemical Physics},
   volume = {145},
   number = {17},
   pages = {170901},
   ISSN = {0021-9606},
   DOI = {10.1063/1.4966192},
   url = {https://doi.org/10.1063/1.4966192},
   year = {2016},
   type = {Journal Article}
}

@article{RN67,
   author = {Brough, David B. and Wheeler, Daniel and Kalidindi, Surya R.},
   title = {Materials Knowledge Systems in Python - A Data Science Framework for Accelerated Development of Hierarchical Materials},
   journal = {Integrating materials and manufacturing innovation},
   volume = {6},
   number = {1},
   pages = {36-53},
   ISSN = {2193-9764
2193-9772},
   DOI = {10.1007/s40192-017-0089-0},
   url = {https://pubmed.ncbi.nlm.nih.gov/28690971
https://www.ncbi.nlm.nih.gov/pmc/articles/PMC5497515/},
   year = {2017},
   type = {Journal Article}
}

@article{RN52,
   author = {Cecen, Ahmet and Fast, Tony and Kalidindi, Surya R.},
   title = {Versatile algorithms for the computation of 2-point spatial correlations in quantifying material structure},
   journal = {Integrating Materials and Manufacturing Innovation},
   volume = {5},
   number = {1},
   pages = {1-15},
   ISSN = {2193-9772},
   DOI = {10.1186/s40192-015-0044-x},
   url = {https://doi.org/10.1186/s40192-015-0044-x},
   year = {2016},
   type = {Journal Article}
}

@article{RN60,
   author = {Cecen, Ahmet and Yabansu, Yuksel C. and Kalidindi, Surya R.},
   title = {A new framework for rotationally invariant two-point spatial correlations in microstructure datasets},
   journal = {Acta Materialia},
   volume = {158},
   pages = {53-64},
   ISSN = {13596454},
   DOI = {10.1016/j.actamat.2018.07.056},
   year = {2018},
   type = {Journal Article}
}

@article{RN75,
   author = {Chen, Pei-En and Xu, Wenxiang and Chawla, N. and Ren, Yi and Jiao and Yang},
   title = {Novel Hierarchical Correlation Functions for Quantitative Representation of Complex Heterogeneous Materials and Microstructural Evolution},
   journal = {arXiv: Materials Science},
   year = {2019},
   type = {Journal Article}
}

@misc{RN7,
   author = {Choudhary, Kamal},
   title = {Jarvis-DFT},
   year = {2014},
   type = {Generic}
}

@article{RN8,
   author = {Choudhary, K. and DeCost, B. and Tavazza, F.},
   title = {Machine learning with force-field inspired descriptors for materials: fast screening and mapping energy landscape},
   journal = {Phys Rev Mater},
   volume = {2},
   number = {8},
   ISSN = {2475-9953 (Print)},
   DOI = {10.1103/physrevmaterials.2.083801},
   url = {https://www.ncbi.nlm.nih.gov/pubmed/32166213},
   year = {2018},
   type = {Journal Article}
}

@article{RN9,
   author = {Choudhary, Kamal and Garrity, Kevin and Tavazza, Francesca},
   title = {Data-driven Discovery of 3D and 2D Thermoelectric Materials},
   journal = {arXiv preprint arXiv:1906.06024},
   year = {2019},
   type = {Journal Article}
}

@article{RN10,
   author = {Choudhary, Kamal and Kalish, Irina and Beams, Ryan and Tavazza, Francesca},
   title = {High-throughput Identification and Characterization of Two-dimensional Materials using Density functional theory},
   journal = {Scientific reports},
   volume = {7},
   number = {1},
   pages = {1-16},
   ISSN = {2045-2322},
   year = {2017},
   type = {Journal Article}
}

@article{RN11,
   author = {Choudhary, Kamal and Tavazza, Francesca},
   title = {Convergence and machine learning predictions of Monkhorst-Pack k-points and plane-wave cut-off in high-throughput DFT calculations},
   journal = {Computational Materials Science},
   volume = {161},
   pages = {300-308},
   ISSN = {0927-0256},
   year = {2019},
   type = {Journal Article}
}

@article{RN12,
   author = {Choudhary, Kamal and Zhang, Qin and Reid, Andrew CE and Chowdhury, Sugata and Van Nguyen, Nhan and Trautt, Zachary and Newrock, Marcus W and Congo, Faical Yannick and Tavazza, Francesca},
   title = {Computational screening of high-performance optoelectronic materials using OptB88vdW and TB-mBJ formalisms},
   journal = {Scientific data},
   volume = {5},
   pages = {180082},
   ISSN = {2052-4463},
   year = {2018},
   type = {Journal Article}
}

@article{RN97,
   author = {Choudhary, Kamal and Zhang, Qin and Reid, Andrew C. E. and Chowdhury, Sugata and Van Nguyen, Nhan and Trautt, Zachary and Newrock, Marcus W. and Congo, Faical Yannick and Tavazza, Francesca},
   title = {Computational screening of high-performance optoelectronic materials using OptB88vdW and TB-mBJ formalisms},
   journal = {Scientific Data},
   volume = {5},
   number = {1},
   pages = {180082},
   ISSN = {2052-4463},
   DOI = {10.1038/sdata.2018.82},
   url = {https://doi.org/10.1038/sdata.2018.82},
   year = {2018},
   type = {Journal Article}
}

@article{RN94,
   author = {Cohn, David A.},
   title = {Neural Network Exploration Using Optimal Experiment Design},
   journal = {Neural Networks},
   volume = {9},
   number = {6},
   pages = {1071-1083},
   ISSN = {0893-6080},
   DOI = {https://doi.org/10.1016/0893-6080(95)00137-9},
   url = {http://www.sciencedirect.com/science/article/pii/0893608095001379},
   year = {1996},
   type = {Journal Article}
}

@article{RN65,
   author = {Debye, P. and Jr., H. R. Anderson and Brumberger, H.},
   title = {Scattering by an Inhomogeneous Solid. II. The Correlation Function and Its Application},
   journal = {Journal of Applied Physics},
   volume = {28},
   number = {6},
   pages = {679-683},
   DOI = {10.1063/1.1722830},
   url = {https://aip.scitation.org/doi/abs/10.1063/1.1722830},
   year = {1957},
   type = {Journal Article}
}

@article{RN19,
   author = {Deml, Ann M. and O’Hayre, Ryan and Wolverton, Chris and Stevanović, Vladan},
   title = {Predicting density functional theory total energies and enthalpies of formation of metal-nonmetal compounds by linear regression},
   journal = {Physical Review B},
   volume = {93},
   number = {8},
   ISSN = {2469-9950
2469-9969},
   DOI = {10.1103/PhysRevB.93.085142},
   year = {2016},
   type = {Journal Article}
}

@article{RN85,
   author = {Duvenaud, David},
   title = {Automatic model construction with Gaussian processes},
   year = {2015},
   type = {Journal Article}
}

@article{RN35,
   author = {Egorova, O. and Hafizi, R. and Woods, D. C. and Day, G. M.},
   title = {Multifidelity Statistical Machine Learning for Molecular Crystal Structure Prediction},
   journal = {J Phys Chem A},
   volume = {124},
   number = {39},
   pages = {8065-8078},
   ISSN = {1520-5215 (Electronic)
1089-5639 (Linking)},
   DOI = {10.1021/acs.jpca.0c05006},
   url = {https://www.ncbi.nlm.nih.gov/pubmed/32881496},
   year = {2020},
   type = {Journal Article}
}

@article{RN20,
   author = {Faber, Felix and Lindmaa, Alexander and von Lilienfeld, O. Anatole and Armiento, Rickard},
   title = {Crystal structure representations for machine learning models of formation energies},
   journal = {International Journal of Quantum Chemistry},
   volume = {115},
   number = {16},
   pages = {1094-1101},
   ISSN = {00207608},
   DOI = {10.1002/qua.24917},
   year = {2015},
   type = {Journal Article}
}

@article{RN68,
   author = {Fast, Tony and Kalidindi, Surya R.},
   title = {Formulation and calibration of higher-order elastic localization relationships using the MKS approach},
   journal = {Acta Materialia},
   volume = {59},
   number = {11},
   pages = {4595-4605},
   ISSN = {1359-6454},
   DOI = {https://doi.org/10.1016/j.actamat.2011.04.005},
   url = {http://www.sciencedirect.com/science/article/pii/S1359645411002473},
   year = {2011},
   type = {Journal Article}
}

@article{RN70,
   author = {Fernandez-Zelaia, Patxi and Yabansu, Yuksel C. and Kalidindi, Surya R.},
   title = {A Comparative Study of the Efficacy of Local/Global and Parametric/Nonparametric Machine Learning Methods for Establishing Structure–Property Linkages in High-Contrast 3D Elastic Composites},
   journal = {Integrating Materials and Manufacturing Innovation},
   volume = {8},
   number = {2},
   pages = {67-81},
   ISSN = {2193-9764
2193-9772},
   DOI = {10.1007/s40192-019-00129-4},
   year = {2019},
   type = {Journal Article}
}

@article{RN74,
   author = {Fullwood, David T. and Niezgoda, Stephen R. and Kalidindi, Surya R.},
   title = {Microstructure reconstructions from 2-point statistics using phase-recovery algorithms},
   journal = {Acta Materialia},
   volume = {56},
   number = {5},
   pages = {942-948},
   ISSN = {1359-6454},
   DOI = {https://doi.org/10.1016/j.actamat.2007.10.044},
   url = {http://www.sciencedirect.com/science/article/pii/S1359645407007458},
   year = {2008},
   type = {Journal Article}
}

@article{RN50,
   author = {Girosi, Federico and Jones, Michael and Poggio, Tomaso},
   title = {Regularization Theory and Neural Networks Architectures},
   journal = {Neural Computation},
   volume = {7},
   number = {2},
   pages = {219-269},
   DOI = {10.1162/neco.1995.7.2.219},
   url = {https://www.mitpressjournals.org/doi/abs/10.1162/neco.1995.7.2.219.},
   year = {1995},
   type = {Journal Article}
}

@article{RN39,
   author = {Gladkikh, Vladislav and Kim, Dong Yeon and Hajibabaei, Amir and Jana, Atanu and Myung, Chang Woo and Kim, Kwang S.},
   title = {Machine Learning for Predicting the Band Gaps of ABX3 Perovskites from Elemental Properties},
   journal = {The Journal of Physical Chemistry C},
   volume = {124},
   number = {16},
   pages = {8905-8918},
   ISSN = {1932-7447
1932-7455},
   DOI = {10.1021/acs.jpcc.9b11768},
   year = {2020},
   type = {Journal Article}
}

@article{RN66,
   author = {Gokhale, A. M. and Tewari, A. and Garmestani, H.},
   title = {Constraints on microstructural two-point correlation functions},
   journal = {Scripta Materialia},
   volume = {53},
   number = {8},
   pages = {989-993},
   ISSN = {1359-6462},
   DOI = {https://doi.org/10.1016/j.scriptamat.2005.06.013},
   url = {http://www.sciencedirect.com/science/article/pii/S1359646205003702},
   year = {2005},
   type = {Journal Article}
}

@article{RN72,
   author = {Gomberg, J. A. and Medford, A. J. and Kalidindi, S. R.},
   title = {Extracting knowledge from molecular mechanics simulations of grain boundaries using machine learning},
   journal = {Acta Materialia},
   volume = {133},
   pages = {100-108},
   DOI = {10.1016/j.actamat.2017.05.009},
   url = {https://www.scopus.com/inward/record.uri?eid=2-s2.0-85019878511&doi=10.1016%2fj.actamat.2017.05.009&partnerID=40&md5=f211f2a5a55cf9c1bdc9b99888866cc5},
   year = {2017},
   type = {Journal Article}
}

@article{RN92,
   author = {Gramacy, Robert and Apley, Daniel},
   title = {Local Gaussian Process Approximation for Large Computer Experiments},
   journal = {Journal of Computational and Graphical Statistics},
   volume = {24},
   DOI = {10.1080/10618600.2014.914442},
   year = {2013},
   type = {Journal Article}
}

@article{RN55,
   author = {Gramacy, Robert B.},
   title = {laGP: Large-Scale Spatial Modeling via Local Approximate Gaussian Processes in R},
   journal = {Journal of Statistical Software; Vol 1, Issue 1 (2016)},
   url = {https://www.jstatsoft.org/v072/i01
http://dx.doi.org/10.18637/jss.v072.i01},
   year = {2016},
   type = {Journal Article}
}

@article{RN91,
   author = {Gramacy, Robert B. and Lee, Herbert K. H.},
   title = {Bayesian Treed Gaussian Process Models With an Application to Computer Modeling},
   journal = {Journal of the American Statistical Association},
   volume = {103},
   number = {483},
   pages = {1119-1130},
   ISSN = {0162-1459},
   DOI = {10.1198/016214508000000689},
   url = {https://doi.org/10.1198/016214508000000689},
   year = {2008},
   type = {Journal Article}
}

@article{RN33,
   author = {Graser, Jake and Kauwe, Steven K. and Sparks, Taylor D.},
   title = {Machine Learning and Energy Minimization Approaches for Crystal Structure Predictions: A Review and New Horizons},
   journal = {Chemistry of Materials},
   volume = {30},
   number = {11},
   pages = {3601-3612},
   ISSN = {0897-4756
1520-5002},
   DOI = {10.1021/acs.chemmater.7b05304},
   year = {2018},
   type = {Journal Article}
}

@inproceedings{RN57,
   author = {Gray, Alexander G and Moore, Andrew W},
   title = {N-body'problems in statistical learning},
   booktitle = {Advances in neural information processing systems},
   pages = {521-527},
   type = {Conference Proceedings}
}

@article{RN64,
   author = {Gupta, Akash and Cecen, Ahmet and Goyal, Sharad and Singh, Amarendra K. and Kalidindi, Surya R.},
   title = {Structure–property linkages using a data science approach: Application to a non-metallic inclusion/steel composite system},
   journal = {Acta Materialia},
   volume = {91},
   pages = {239-254},
   ISSN = {1359-6454},
   DOI = {https://doi.org/10.1016/j.actamat.2015.02.045},
   url = {http://www.sciencedirect.com/science/article/pii/S1359645415001603},
   year = {2015},
   type = {Journal Article}
}

@article{RN13,
   author = {Haastrup, Sten and Strange, Mikkel and Pandey, Mohnish and Deilmann, Thorsten and Schmidt, Per S. and Hinsche, Nicki F. and Gjerding, Morten N. and Torelli, Daniele and Larsen, Peter M. and Riis-Jensen, Anders C. and Gath, Jakob and Jacobsen, Karsten W. and Jørgen Mortensen, Jens and Olsen, Thomas and Thygesen, Kristian S.},
   title = {The Computational 2D Materials Database: high-throughput modeling and discovery of atomically thin crystals},
   journal = {2D Materials},
   volume = {5},
   number = {4},
   pages = {042002},
   ISSN = {2053-1583},
   DOI = {10.1088/2053-1583/aacfc1},
   url = {http://dx.doi.org/10.1088/2053-1583/aacfc1},
   year = {2018},
   type = {Journal Article}
}

@article{RN37,
   author = {Hansen, Katja and Biegler, Franziska and Ramakrishnan, Raghunathan and Pronobis, Wiktor and von Lilienfeld, O. Anatole and Müller, Klaus-Robert and Tkatchenko, Alexandre},
   title = {Machine Learning Predictions of Molecular Properties: Accurate Many-Body Potentials and Nonlocality in Chemical Space},
   journal = {The Journal of Physical Chemistry Letters},
   volume = {6},
   number = {12},
   pages = {2326-2331},
   DOI = {10.1021/acs.jpclett.5b00831},
   url = {https://doi.org/10.1021/acs.jpclett.5b00831},
   year = {2015},
   type = {Journal Article}
}

@article{RN2,
   author = {Hohenberg, P. and Kohn, W.},
   title = {Inhomogeneous Electron Gas},
   journal = {Physical Review},
   volume = {136},
   number = {3B},
   pages = {B864-B871},
   DOI = {10.1103/PhysRev.136.B864},
   url = {https://link.aps.org/doi/10.1103/PhysRev.136.B864},
   year = {1964},
   type = {Journal Article}
}

@article{RN42,
   author = {Honrao, Shreyas and Anthonio, Bryan E. and Ramanathan, Rohit and Gabriel, Joshua J. and Hennig, Richard G.},
   title = {Machine learning of ab-initio energy landscapes for crystal structure predictions},
   journal = {Computational Materials Science},
   volume = {158},
   pages = {414-419},
   ISSN = {0927-0256},
   DOI = {https://doi.org/10.1016/j.commatsci.2018.08.041},
   url = {http://www.sciencedirect.com/science/article/pii/S092702561830569X},
   year = {2019},
   type = {Journal Article}
}

@article{RN43,
   author = {Honrao, Shreyas J. and Xie, Stephen R. and Hennig, Richard G.},
   title = {Augmenting machine learning of energy landscapes with local structural information},
   journal = {Journal of Applied Physics},
   volume = {128},
   number = {8},
   pages = {085101},
   ISSN = {0021-8979},
   DOI = {10.1063/5.0012407},
   url = {https://doi.org/10.1063/5.0012407},
   year = {2020},
   type = {Journal Article}
}

@article{RN26,
   author = {Huo, Haoyan and Rupp, Matthias},
   title = {Unified Representation for Machine Learning of Molecules and Crystals},
   year = {2017},
   type = {Journal Article}
}

@article{RN16,
   author = {Jain, Anubhav and Ong, Shyue and Hautier, Geoffroy and Chen, Wei and Richards, William and Dacek, Stephen and Cholia, Shreyas and Gunter, Dan and Skinner, David and Ceder, Gerbrand and Persson, Kristin},
   title = {Commentary: The Materials Project: A materials genome approach to accelerating materials innovation},
   journal = {APL Materials},
   volume = {1},
   pages = {011002},
   DOI = {10.1063/1.4812323},
   year = {2013},
   type = {Journal Article}
}

@article{RN40,
   author = {Jha, Dipendra and Ward, Logan and Paul, Arindam and Liao, Wei-keng and Choudhary, Alok and Wolverton, Chris and Agrawal, Ankit},
   title = {ElemNet: Deep Learning the Chemistry of Materials From Only Elemental Composition},
   journal = {Scientific Reports},
   volume = {8},
   number = {1},
   pages = {17593},
   ISSN = {2045-2322},
   DOI = {10.1038/s41598-018-35934-y},
   url = {https://doi.org/10.1038/s41598-018-35934-y},
   year = {2018},
   type = {Journal Article}
}

@article{RN53,
   author = {Kajita, Seiji and Ohba, Nobuko and Jinnouchi, Ryosuke and Asahi, Ryoji},
   title = {A Universal 3D Voxel Descriptor for Solid-State Material Informatics with Deep Convolutional Neural Networks},
   journal = {Scientific Reports},
   volume = {7},
   number = {1},
   pages = {16991},
   ISSN = {2045-2322},
   DOI = {10.1038/s41598-017-17299-w},
   url = {https://doi.org/10.1038/s41598-017-17299-w},
   year = {2017},
   type = {Journal Article}
}

@book{RN5,
   author = {Kalidindi, Surya},
   title = {Hierarchical Materials Informatics: Novel Analytics for Materials Data},
   pages = {1-219},
   year = {2015},
   type = {Book}
}

@article{RN69,
   author = {Kalidindi, Surya R.},
   title = {Computationally Efficient, Fully Coupled Multiscale Modeling of Materials Phenomena Using Calibrated Localization Linkages},
   journal = {ISRN Materials Science},
   volume = {2012},
   pages = {305692},
   ISSN = {xxxx-xxxx},
   DOI = {10.5402/2012/305692},
   url = {https://doi.org/10.5402/2012/305692},
   year = {2012},
   type = {Journal Article}
}

@article{RN71,
   author = {Kalidindi, S. R.},
   title = {A Bayesian framework for materials knowledge systems},
   journal = {MRS Communications},
   volume = {9},
   number = {2},
   pages = {518-531},
   DOI = {10.1557/mrc.2019.56},
   url = {https://www.scopus.com/inward/record.uri?eid=2-s2.0-85065401757&doi=10.1557%2fmrc.2019.56&partnerID=40&md5=bae53ebe5353b3b3daa4961f6384804f},
   year = {2019},
   type = {Journal Article}
}

@article{RN95,
   author = {Kalidindi, S. R. and Gomberg, J. A. and Trautt, Z. T. and Becker, C. A.},
   title = {Application of data science tools to quantify and distinguish between structures and models in molecular dynamics datasets},
   journal = {Nanotechnology},
   volume = {26},
   number = {34},
   pages = {344006},
   ISSN = {1361-6528 (Electronic)
0957-4484 (Linking)},
   DOI = {10.1088/0957-4484/26/34/344006},
   url = {https://www.ncbi.nlm.nih.gov/pubmed/26235174},
   year = {2015},
   type = {Journal Article}
}

@article{RN56,
   author = {Kalidindi, Surya R. and Niezgoda, Stephen R. and Salem, Ayman A.},
   title = {Microstructure informatics using higher-order statistics and efficient data-mining protocols},
   journal = {JOM},
   volume = {63},
   number = {4},
   pages = {34-41},
   ISSN = {1543-1851},
   DOI = {10.1007/s11837-011-0057-7},
   url = {https://doi.org/10.1007/s11837-011-0057-7},
   year = {2011},
   type = {Journal Article}
}

@article{RN44,
   author = {Karamad, Mohammadreza and Magar, Rishikesh and Shi, Yuting and Siahrostami, Samira and Gates, Ian D. and Barati Farimani, Amir},
   title = {Orbital graph convolutional neural network for material property prediction},
   journal = {Physical Review Materials},
   volume = {4},
   number = {9},
   pages = {093801},
   DOI = {10.1103/PhysRevMaterials.4.093801},
   url = {https://link.aps.org/doi/10.1103/PhysRevMaterials.4.093801},
   year = {2020},
   type = {Journal Article}
}

@misc{RN101,
   author = {Kaundinya, P. R. and Choudhary, K. and Kalidindi, S. R.},
   title = {https://materialsdata.nist.gov/handle/11256/994},
   url = {https://materialsdata.nist.gov/handle/11256/994},
   type = {Dataset}
}

@article{RN99,
   author = {Kingma, Diederik and Ba, Jimmy},
   title = {Adam: A Method for Stochastic Optimization},
   journal = {International Conference on Learning Representations},
   year = {2014},
   type = {Journal Article}
}

@article{RN47,
   author = {Kipf, Thomas N and Welling, Max},
   title = {Semi-supervised classification with graph convolutional networks},
   journal = {arXiv preprint arXiv:1609.02907},
   year = {2016},
   type = {Journal Article}
}

@article{RN6,
   author = {Kirklin, Scott and Saal, James E. and Meredig, Bryce and Thompson, Alex and Doak, Jeff W. and Aykol, Muratahan and Rühl, Stephan and Wolverton, Chris},
   title = {The Open Quantum Materials Database (OQMD): assessing the accuracy of DFT formation energies},
   journal = {npj Computational Materials},
   volume = {1},
   number = {1},
   ISSN = {2057-3960},
   DOI = {10.1038/npjcompumats.2015.10},
   year = {2015},
   type = {Journal Article}
}

@article{RN1,
   author = {Kohn, W. and Sham, L. J.},
   title = {Self-Consistent Equations Including Exchange and Correlation Effects},
   journal = {Physical Review},
   volume = {140},
   number = {4A},
   pages = {A1133-A1138},
   ISSN = {0031-899X},
   DOI = {10.1103/PhysRev.140.A1133},
   year = {1965},
   type = {Journal Article}
}

@article{RN81,
   author = {Krige, D.G.},
   title = {A statistical approach to some basic mine valuation problems on the Witwatersrand},
   journal = {Journal of the Southern African Institute of Mining and Metallurgy},
   volume = {52},
   number = {6},
   pages = {119-139},
   ISSN = {0038-223X},
   url = {https://journals.co.za/content/saimm/52/6/AJA0038223X_4792},
   year = {1951},
   type = {Journal Article}
}

@article{RN17,
   author = {Lee, Joohwi and Seko, Atsuto and Shitara, Kazuki and Nakayama, Keita and Tanaka, Isao},
   title = {Prediction model of band gap for inorganic compounds by combination of density functional theory calculations and machine learning techniques},
   journal = {Physical Review B},
   volume = {93},
   number = {11},
   ISSN = {2469-9950
2469-9969},
   DOI = {10.1103/PhysRevB.93.115104},
   year = {2016},
   type = {Journal Article}
}

@article{RN86,
   author = {Leithead, W. E. and Zhang, Yunong},
   title = {O(N 2)-Operation Approximation of Covariance Matrix Inverse in Gaussian Process Regression Based on Quasi-Newton BFGS Method},
   journal = {Communications in Statistics - Simulation and Computation},
   volume = {36},
   number = {2},
   pages = {367-380},
   ISSN = {0361-0918},
   DOI = {10.1080/03610910601161298},
   url = {https://www.tandfonline.com/doi/abs/10.1080/03610910601161298},
   year = {2007},
   type = {Journal Article}
}

@inproceedings{RN89,
   author = {McIntire M. and Ratner, D. and Ermon, S.},
   title = {Sparse Gaussian processes for Bayesian optimization},
   pages = {517–526},
   year = {2016},
   booktitle = {Proceedings of the Thirty-Second Conference on Uncertainty in Artificial Intelligence},
   type = {Conference Paper}
}

@article{RN34,
   author = {Meredig, B. and Agrawal, A. and Kirklin, S. and Saal, J. E. and Doak, J. W. and Thompson, A. and Zhang, K. and Choudhary, A. and Wolverton, C.},
   title = {Combinatorial screening for new materials in unconstrained composition space with machine learning},
   journal = {Physical Review B},
   volume = {89},
   number = {9},
   pages = {094104},
   DOI = {10.1103/PhysRevB.89.094104},
   url = {https://link.aps.org/doi/10.1103/PhysRevB.89.094104},
   year = {2014},
   type = {Journal Article}
}

@article{RN46,
   author = {Micheli, A.},
   title = {Neural Network for Graphs: A Contextual Constructive Approach},
   journal = {IEEE Transactions on Neural Networks},
   volume = {20},
   number = {3},
   pages = {498-511},
   ISSN = {1941-0093},
   DOI = {10.1109/TNN.2008.2010350},
   year = {2009},
   type = {Journal Article}
}

@inbook{RN27,
   author = {Montavon, Grégoire and Hansen, Katja and Fazli, Siamac and Rupp, Matthias and Biegler, Franziska and Ziehe, Andreas and Tkatchenko, Alexandre and von Lilienfeld, Anatole and Müller, Klaus-Robert},
   title = {Learning Invariant Representations of Molecules for Atomization Energy Prediction},
   volume = {1},
   pages = {449-457},
   year = {2012},
   type = {Book Section}
}

@inproceedings{RN58,
   author = {Moore, Andrew W. and Connolly, Andy J. and Genovese, Chris and Gray, Alex and Grone, Larry and Kanidoris II, Nick and Nichol, Robert C. and Schneider, Jeff and Szalay, Alex S. and Szapudi, Istvan and Wasserman, Larry},
   title = {Fast Algorithms and Efficient Statistics: N-Point Correlation Functions},
   series = {Mining the Sky},
   publisher = {Springer Berlin Heidelberg},
   pages = {71-82},
   ISBN = {978-3-540-44665-1},
   type = {Conference Proceedings}
}

@article{RN88,
   author = {Nelder, J. A. and Mead, R.},
   title = {A Simplex Method for Function Minimization},
   journal = {The Computer Journal},
   volume = {7},
   number = {4},
   pages = {308-313},
   ISSN = {0010-4620},
   DOI = {10.1093/comjnl/7.4.308},
   url = {https://doi.org/10.1093/comjnl/7.4.308},
   year = {1965},
   type = {Journal Article}
}

@article{RN98,
   author = {Nie, Feiping and Zhanxuan, Hu and Li, Xuelong},
   title = {An investigation for loss functions widely used in machine learning},
   journal = {Communications in Information and Systems},
   volume = {18},
   pages = {37-52},
   DOI = {10.4310/CIS.2018.v18.n1.a2},
   year = {2018},
   type = {Journal Article}
}

@article{RN61,
   author = {Niezgoda, S. R. and Fullwood, D. T. and Kalidindi, S. R.},
   title = {Delineation of the space of 2-point correlations in a composite material system},
   journal = {Acta Materialia},
   volume = {56},
   number = {18},
   pages = {5285-5292},
   ISSN = {1359-6454},
   DOI = {https://doi.org/10.1016/j.actamat.2008.07.005},
   url = {http://www.sciencedirect.com/science/article/pii/S1359645408004886},
   year = {2008},
   type = {Journal Article}
}

@article{RN77,
   author = {Niezgoda, Stephen R. and Turner, David M. and Fullwood, David T. and Kalidindi, Surya R.},
   title = {Optimized structure based representative volume element sets reflecting the ensemble-averaged 2-point statistics},
   journal = {Acta Materialia},
   volume = {58},
   number = {13},
   pages = {4432-4445},
   ISSN = {13596454},
   DOI = {10.1016/j.actamat.2010.04.041},
   year = {2010},
   type = {Journal Article}
}

@article{RN87,
   author = {Nocedal, Jorge},
   title = {Updating Quasi-Newton Matrices with Limited Storage},
   journal = {Mathematics of Computation},
   volume = {35},
   number = {151},
   pages = {773-782},
   ISSN = {00255718, 10886842},
   DOI = {10.2307/2006193},
   url = {www.jstor.org/stable/2006193},
   year = {1980},
   type = {Journal Article}
}

@book{RN78,
   author = {Nwankpa, Chigozie and Ijomah, Winifred and Gachagan, Anthony and Marshall, Stephen},
   title = {Activation Functions: Comparison of trends in Practice and Research for Deep Learning},
   year = {2018},
   type = {Book}
}

@article{RN45,
   author = {Park, Cheol Woo and Wolverton, Chris},
   title = {Developing an improved crystal graph convolutional neural network framework for accelerated materials discovery},
   journal = {Physical Review Materials},
   volume = {4},
   number = {6},
   pages = {063801},
   DOI = {10.1103/PhysRevMaterials.4.063801},
   url = {https://link.aps.org/doi/10.1103/PhysRevMaterials.4.063801},
   year = {2020},
   type = {Journal Article}
}

@article{RN79,
   author = {Paszke, Adam and Gross, S. and Massa, Francisco and Lerer, A. and Bradbury, J. and Chanan, G. and Killeen, T. and Lin, Z. and Gimelshein, N. and Antiga, L. and Desmaison, Alban and Köpf, Andreas and Yang, E. and DeVito, Zach and Raison, Martin and Tejani, Alykhan and Chilamkurthy, Sasank and Steiner, B. and Fang, Lu and Bai, Junjie and Chintala, Soumith},
   title = {PyTorch: An Imperative Style, High-Performance Deep Learning Library},
   journal = {ArXiv},
   volume = {abs/1912.01703},
   year = {2019},
   type = {Journal Article}
}

@article{RN73,
   author = {Paulson, N. H. and Priddy, M. W. and McDowell, D. L. and Kalidindi, S. R.},
   title = {Reduced-order structure-property linkages for polycrystalline microstructures based on 2-point statistics},
   journal = {Acta Materialia},
   volume = {129},
   pages = {428-438},
   DOI = {10.1016/j.actamat.2017.03.009},
   url = {https://www.scopus.com/inward/record.uri?eid=2-s2.0-85015628206&doi=10.1016%2fj.actamat.2017.03.009&partnerID=40&md5=278cca3594a6a53152b9fa13aecb86d3},
   year = {2017},
   type = {Journal Article}
}

@article{RN18,
   author = {Pilania, G. and Mannodi-Kanakkithodi, A. and Uberuaga, B. P. and Ramprasad, R. and Gubernatis, J. E. and Lookman, T.},
   title = {Machine learning bandgaps of double perovskites},
   journal = {Scientific Reports},
   volume = {6},
   number = {1},
   pages = {19375},
   ISSN = {2045-2322},
   DOI = {10.1038/srep19375},
   url = {https://doi.org/10.1038/srep19375},
   year = {2016},
   type = {Journal Article}
}

@article{RN36,
   author = {Podryabinkin, Evgeny V. and Tikhonov, Evgeny V. and Shapeev, Alexander V. and Oganov, Artem R.},
   title = {Accelerating crystal structure prediction by machine-learning interatomic potentials with active learning},
   journal = {Physical Review B},
   volume = {99},
   number = {6},
   ISSN = {2469-9950
2469-9969},
   DOI = {10.1103/PhysRevB.99.064114},
   year = {2019},
   type = {Journal Article}
}

@inbook{RN48,
   author = {Prechelt, Lutz},
   title = {Early Stopping — But When?},
   booktitle = {Neural Networks: Tricks of the Trade: Second Edition},
   editor = {Montavon, Grégoire and Orr, Geneviève B. and Müller, Klaus-Robert},
   publisher = {Springer Berlin Heidelberg},
   address = {Berlin, Heidelberg},
   pages = {53-67},
   ISBN = {978-3-642-35289-8},
   DOI = {10.1007/978-3-642-35289-8_5},
   url = {https://doi.org/10.1007/978-3-642-35289-8_5},
   year = {2012},
   type = {Book Section}
}

@article{RN90,
   author = {Quiñonero-Candela, Joaquin and Rasmussen, Carl Edward},
   title = {A Unifying View of Sparse Approximate Gaussian Process Regression},
   journal = {J. Mach. Learn. Res.},
   volume = {6},
   pages = {1939–1959},
   ISSN = {1532-4435},
   year = {2005},
   type = {Journal Article}
}

@article{RN15,
   author = {Ramakrishnan, Raghunathan and Dral, Pavlo O. and Rupp, Matthias and von Lilienfeld, O. Anatole},
   title = {Quantum chemistry structures and properties of 134 kilo molecules},
   journal = {Scientific Data},
   volume = {1},
   number = {1},
   pages = {140022},
   ISSN = {2052-4463},
   DOI = {10.1038/sdata.2014.22},
   url = {https://doi.org/10.1038/sdata.2014.22},
   year = {2014},
   type = {Journal Article}
}

@article{RN29,
   author = {Ramakrishnan, R. and Dral, P. O. and Rupp, M. and von Lilienfeld, O. A.},
   title = {Big Data Meets Quantum Chemistry Approximations: The Delta-Machine Learning Approach},
   journal = {J Chem Theory Comput},
   volume = {11},
   number = {5},
   pages = {2087-96},
   ISSN = {1549-9626 (Electronic)
1549-9618 (Linking)},
   DOI = {10.1021/acs.jctc.5b00099},
   url = {https://www.ncbi.nlm.nih.gov/pubmed/26574412},
   year = {2015},
   type = {Journal Article}
}

@book{RN51,
   author = {Rasmussen, Carl Edward and Williams, Christopher K. I.},
   title = {Gaussian processes for machine learning},
   publisher = {MIT Press},
   address = {Cambridge, Mass.},
   series = {Adaptive computation and machine learning},
   pages = {xviii, 248 p.},
   ISBN = {026218253X},
   url = {Table of contents only http://www.loc.gov/catdir/toc/fy0614/2005053433.html},
   year = {2006},
   type = {Book}
}

@article{RN14,
   author = {Ruddigkeit, Lars and van Deursen, Ruud and Blum, Lorenz C. and Reymond, Jean-Louis},
   title = {Enumeration of 166 Billion Organic Small Molecules in the Chemical Universe Database GDB-17},
   journal = {Journal of Chemical Information and Modeling},
   volume = {52},
   number = {11},
   pages = {2864-2875},
   ISSN = {1549-9596},
   DOI = {10.1021/ci300415d},
   url = {https://doi.org/10.1021/ci300415d},
   year = {2012},
   type = {Journal Article}
}

@article{RN25,
   author = {Rupp, M. and Tkatchenko, A. and Muller, K. R. and von Lilienfeld, O. A.},
   title = {Fast and accurate modeling of molecular atomization energies with machine learning},
   journal = {Phys Rev Lett},
   volume = {108},
   number = {5},
   pages = {058301},
   ISSN = {1079-7114 (Electronic)
0031-9007 (Linking)},
   DOI = {10.1103/PhysRevLett.108.058301},
   url = {https://www.ncbi.nlm.nih.gov/pubmed/22400967},
   year = {2012},
   type = {Journal Article}
}

@article{RN32,
   author = {Ryan, K. and Lengyel, J. and Shatruk, M.},
   title = {Crystal Structure Prediction via Deep Learning},
   journal = {J Am Chem Soc},
   volume = {140},
   number = {32},
   pages = {10158-10168},
   ISSN = {1520-5126 (Electronic)
0002-7863 (Linking)},
   DOI = {10.1021/jacs.8b03913},
   url = {https://www.ncbi.nlm.nih.gov/pubmed/29874459},
   year = {2018},
   type = {Journal Article}
}

@article{RN82,
   author = {Schulz, Eric and Speekenbrink, Maarten and Krause, Andreas},
   title = {A tutorial on Gaussian process regression: Modelling, exploring, and exploiting functions},
   journal = {Journal of Mathematical Psychology},
   volume = {85},
   pages = {1-16},
   ISSN = {00222496},
   DOI = {10.1016/j.jmp.2018.03.001},
   year = {2018},
   type = {Journal Article}
}

@article{RN41,
   author = {Schütt, K. T. and Glawe, H. and Brockherde, F. and Sanna, A. and Müller, K. R. and Gross, E. K. U.},
   title = {How to represent crystal structures for machine learning: Towards fast prediction of electronic properties},
   journal = {Physical Review B},
   volume = {89},
   number = {20},
   ISSN = {1098-0121
1550-235X},
   DOI = {10.1103/PhysRevB.89.205118},
   year = {2014},
   type = {Journal Article}
}

@article{RN49,
   author = {Srivastava, Nitish and Hinton, Geoffrey and Krizhevsky, Alex and Sutskever, Ilya and Salakhutdinov, Ruslan},
   title = {Dropout: a simple way to prevent neural networks from overfitting},
   journal = {J. Mach. Learn. Res.},
   volume = {15},
   number = {1},
   pages = {1929–1958},
   ISSN = {1532-4435},
   year = {2014},
   type = {Journal Article}
}

@article{RN59,
   author = {Torquato, Salvatore and Haslach Jr, HW},
   title = {Random heterogeneous materials: microstructure and macroscopic properties},
   journal = {Appl. Mech. Rev.},
   volume = {55},
   number = {4},
   pages = {B62-B63},
   ISSN = {0003-6900},
   year = {2002},
   type = {Journal Article}
}

@article{RN93,
   author = {Vecchia, A. V.},
   title = {Estimation and Model Identification for Continuous Spatial Processes},
   journal = {Journal of the Royal Statistical Society. Series B (Methodological)},
   volume = {50},
   number = {2},
   pages = {297-312},
   ISSN = {00359246},
   url = {www.jstor.org/stable/2345768},
   year = {1988},
   type = {Journal Article}
}

@article{RN28,
   author = {Ward, Logan and Blaiszik, Ben and Foster, Ian and Assary, Rajeev S. and Narayanan, Badri and Curtiss, Larry},
   title = {Machine learning prediction of accurate atomization energies of organic molecules from low-fidelity quantum chemical calculations},
   journal = {MRS Communications},
   volume = {9},
   number = {3},
   pages = {891-899},
   ISSN = {2159-6859},
   DOI = {10.1557/mrc.2019.107},
   year = {2019},
   type = {Journal Article}
}

@article{RN23,
   author = {Ward, Logan and Liu, Ruoqian and Krishna, Amar and Hegde, Vinay I. and Agrawal, Ankit and Choudhary, Alok and Wolverton, Chris},
   title = {Including crystal structure attributes in machine learning models of formation energies via Voronoi tessellations},
   journal = {Physical Review B},
   volume = {96},
   number = {2},
   ISSN = {2469-9950
2469-9969},
   DOI = {10.1103/PhysRevB.96.024104},
   year = {2017},
   type = {Journal Article}
}

@article{RN100,
   author = {Weaver, Brian and Williams, Brian and Anderson-Cook, Christine and Higdon, David},
   title = {Computational Enhancements to Bayesian Design of Experiments Using Gaussian Processes},
   journal = {Bayesian Analysis},
   volume = {11},
   DOI = {10.1214/15-BA945},
   year = {2015},
   type = {Journal Article}
}

@article{RN30,
   author = {Wilkins, David M. and Grisafi, Andrea and Yang, Yang and Lao, Ka Un and DiStasio, Robert A. and Ceriotti, Michele},
   title = {Accurate molecular polarizabilities with coupled cluster theory and machine learning},
   journal = {Proceedings of the National Academy of Sciences},
   volume = {116},
   number = {9},
   pages = {3401},
   DOI = {10.1073/pnas.1816132116},
   url = {http://www.pnas.org/content/116/9/3401.abstract},
   year = {2019},
   type = {Journal Article}
}

@article{RN62,
   author = {Wolf, W. P.},
   title = {The Ising model and real magnetic materials},
   journal = {Brazilian Journal of Physics},
   volume = {30},
   pages = {794-810},
   ISSN = {0103-9733},
   url = {http://www.scielo.br/scielo.php?script=sci_arttext&pid=S0103-97332000000400030&nrm=iso},
   year = {2000},
   type = {Journal Article}
}

@article{RN63,
   author = {Wu, F. Y.},
   title = {The Potts model},
   journal = {Reviews of Modern Physics},
   volume = {54},
   number = {1},
   pages = {235-268},
   DOI = {10.1103/RevModPhys.54.235},
   url = {https://link.aps.org/doi/10.1103/RevModPhys.54.235},
   year = {1982},
   type = {Journal Article}
}

@article{RN24,
   author = {Xie, Tian and Grossman, Jeffrey C.},
   title = {Crystal Graph Convolutional Neural Networks for an Accurate and Interpretable Prediction of Material Properties},
   journal = {Physical Review Letters},
   volume = {120},
   number = {14},
   pages = {145301},
   DOI = {10.1103/PhysRevLett.120.145301},
   url = {https://link.aps.org/doi/10.1103/PhysRevLett.120.145301},
   year = {2018},
   type = {Journal Article}
}

@article{RN83,
   author = {Yabansu, Y. C. and Iskakov, A. and Kapustina, A. and Rajagopalan, S. and Kalidindi, S. R.},
   title = {Application of Gaussian process regression models for capturing the evolution of microstructure statistics in aging of nickel-based superalloys},
   journal = {Acta Materialia},
   volume = {178},
   pages = {45-58},
   DOI = {10.1016/j.actamat.2019.07.048},
   url = {https://www.scopus.com/inward/record.uri?eid=2-s2.0-85070237889&doi=10.1016%2fj.actamat.2019.07.048&partnerID=40&md5=6b2960fd74795cdf44e39adf703822dd},
   year = {2019},
   type = {Journal Article}
}

@article{RN96,
   author = {Yabansu, Yuksel C. and Kalidindi, Surya R.},
   title = {Representation and calibration of elastic localization kernels for a broad class of cubic polycrystals},
   journal = {Acta Materialia},
   volume = {94},
   pages = {26-35},
   ISSN = {13596454},
   DOI = {10.1016/j.actamat.2015.04.049},
   year = {2015},
   type = {Journal Article}
}

@article{RN84,
   author = {Yabansu, Y. C. and Rehn, V. and Hötzer, J. and Nestler, B. and Kalidindi, S. R.},
   title = {Application of Gaussian process autoregressive models for capturing the time evolution of microstructure statistics from phase-field simulations for sintering of polycrystalline ceramics},
   journal = {Modelling and Simulation in Materials Science and Engineering},
   volume = {27},
   number = {8},
   DOI = {10.1088/1361-651X/ab413e},
   url = {https://www.scopus.com/inward/record.uri?eid=2-s2.0-85081614709&doi=10.1088%2f1361-651X%2fab413e&partnerID=40&md5=313ad8df47a568a7f7b57a7363b92cac},
   year = {2019},
   type = {Journal Article}
}

@article{RN21,
   author = {Ye, W. and Chen, C. and Wang, Z. and Chu, I. H. and Ong, S. P.},
   title = {Deep neural networks for accurate predictions of crystal stability},
   journal = {Nat Commun},
   volume = {9},
   number = {1},
   pages = {3800},
   ISSN = {2041-1723 (Electronic)
2041-1723 (Linking)},
   DOI = {10.1038/s41467-018-06322-x},
   url = {https://www.ncbi.nlm.nih.gov/pubmed/30228262},
   year = {2018},
   type = {Journal Article}
}

@article{RN76,
   author = {Yucel, Berkay and Yucel, Sezen and Ray, Arunim and Duprez, Lode and Kalidindi, Surya R.},
   title = {Mining the Correlations Between Optical Micrographs and Mechanical Properties of Cold-Rolled HSLA Steels Using Machine Learning Approaches},
   journal = {Integrating Materials and Manufacturing Innovation},
   volume = {9},
   number = {3},
   pages = {240-256},
   ISSN = {2193-9764
2193-9772},
   DOI = {10.1007/s40192-020-00183-3},
   year = {2020},
   type = {Journal Article}
}

@article{RN54,
   author = {Zhao, Yong and Yuan, Kunpeng and Liu, Yinqiao and Louis, Steph-Yves and Hu, Ming and Hu, Jianjun},
   title = {Predicting Elastic Properties of Materials from Electronic Charge Density Using 3D Deep Convolutional Neural Networks},
   journal = {The Journal of Physical Chemistry C},
   volume = {124},
   number = {31},
   pages = {17262-17273},
   ISSN = {1932-7447},
   DOI = {10.1021/acs.jpcc.0c02348},
   url = {https://doi.org/10.1021/acs.jpcc.0c02348},
   year = {2020},
   type = {Journal Article}
}

@article{RN31,
   author = {Ziletti, A. and Kumar, D. and Scheffler, M. and Ghiringhelli, L. M.},
   title = {Insightful classification of crystal structures using deep learning},
   journal = {Nat Commun},
   volume = {9},
   number = {1},
   pages = {2775},
   ISSN = {2041-1723 (Electronic)
2041-1723 (Linking)},
   DOI = {10.1038/s41467-018-05169-6},
   url = {https://www.ncbi.nlm.nih.gov/pubmed/30018362},
   year = {2018},
   type = {Journal Article}
}

@article{RN3,
   author = {Zunger, Alex},
   title = {Inverse design in search of materials with target functionalities},
   journal = {Nature Reviews Chemistry},
   volume = {2},
   number = {4},
   pages = {0121},
   ISSN = {2397-3358},
   DOI = {10.1038/s41570-018-0121},
   url = {https://doi.org/10.1038/s41570-018-0121},
   year = {2018},
   type = {Journal Article}
}
	
\end{document}